\documentclass[12pt]{article}
\usepackage{epsfig}
\usepackage{amsmath}
\usepackage{hhline}
\usepackage{amssymb}
\usepackage{times}

\newlength{\dinwidth}
\newlength{\dinmargin}
\newcommand{\lsim}{\raisebox{-1.5mm}{$\:\stackrel{\textstyle{<}}{\textstyle{\sim}}\:$}}
\newcommand{\gsim}{\raisebox{-0.5mm}{$\stackrel{>}{\scriptstyle{\sim}}$}}

\newcommand{\cm}{\mbox{\rm ~cm}}
\def\GeV{\hbox{$\;\hbox{\rm GeV}$}}
\def\MeV{\hbox{$\;\hbox{\rm MeV}$}}

\newcommand{\picob}{\mbox{{\rm ~pb}~}}

\begin{document}
\begin{titlepage}
%
%
\begin{flushleft}
DESY-99-081  \hfill  ISSN0418-9833 \\
June 1999
\end{flushleft}


\vspace*{1.5cm}

\begin{center}
\begin{Large}
\boldmath
\bf{A Search for Leptoquark Bosons and Lepton Flavor Violation 
    in $e^+ p$ Collisions at HERA \\ }
\unboldmath
 
\vspace*{1.2cm}
H1 Collaboration \\
\end{Large}
 
\vspace*{0.8cm}
 
\end{center}
  
\begin{abstract}
\noindent

A search for new bosons possessing couplings to lepton-quark pairs is 
performed in the H1 experiment at HERA using 1994 to 1997 data 
corresponding to an integrated luminosity of $37 \picob^{-1}$.
First generation leptoquarks (LQs) are searched in very high $Q^2$ neutral 
(NC) and charged (CC) current data samples.
The measurements are compared to Standard Model (SM) expectations from 
deep-inelastic scattering (DIS). 
A deviation in the $Q^2$ spectrum previously observed in the 
1994 to 1996 dataset at $Q^2 \, \gsim \, 15000 \GeV^2$ remains, though 
with less significance. 
This deviation corresponded to a clustering in the invariant mass spectrum at
$M \simeq 200 \GeV$ which is not observed with the 1997 dataset alone.
The NC DIS data is used to constrain the Yukawa couplings $\lambda$ of 
first generation scalar and vector LQs in the Buchm\"uller-R\"uckl-Wyler 
effective model. 
Scalar LQs are excluded for masses up to $275 \GeV$ for a coupling
of electromagnetic strength, $\lambda = 0.3$.
A sensitivity to coupling values $\lsim 1$ is established for masses up
to $400 \GeV$ for any LQ type.
The NC and CC DIS data are combined to constrain $\lambda$ for arbitrary 
branching ratios of the LQ into $eq$ in a generic model.
For a decay branching ratio into $e^+ u$ pairs as small as $10 \%$, LQ
masses up to $260 \GeV$ are ruled out for $\lambda = 0.3$.
LQs possessing couplings to mixed fermion generations,
which could lead to signals of lepton flavor violation (LFV),
are searched in events with a high transverse momentum $\mu$ or $\tau$.
No $\mu + X$ or $\tau + X$ event candidate is found that is compatible
with LQ kinematics.
Constraints are set on the Yukawa coupling involving the $\mu$ and $\tau$
lepton in a yet unexplored mass range.
\end{abstract} 

\vspace*{0.7cm}
\begin{center}
{\small{Dedicated to the memory of our colleague and 
        friend Marc~David, deceased 23.01.99.}}
\end{center}

\vspace*{0.3cm}
\begin{center}
Submitted to The European Physical Journal C
\end{center}
 
\vfill

\end{titlepage}

 
%
 C.~Adloff$^{33}$,                
 V.~Andreev$^{24}$,               
 B.~Andrieu$^{27}$,               
 V.~Arkadov$^{34}$,               
 A.~Astvatsatourov$^{34}$,        
 I.~Ayyaz$^{28}$,                 
 A.~Babaev$^{23}$,                
 J.~B\"ahr$^{34}$,                
 P.~Baranov$^{24}$,               
 E.~Barrelet$^{28}$
% ...

 \noindent $ ^1$ I. Physikalisches Institut der RWTH, Aachen, Germany$^a$ \\
% ...
$ ^{23}$ Institute for Theoretical and Experimental Physics,
          Moscow, Russia \\
 $ ^{24}$ Lebedev Physical Institute, Moscow, Russia$^{f,k}$ \\
% ...
$ ^{27}$ LPNHE, \'{E}cole Polytechnique, IN2P3-CNRS, Palaiseau, France \\
 $ ^{28}$ LPNHE, Universit\'{e}s Paris VI and VII, IN2P3-CNRS,
          Paris, France \\
% ...
$ ^{33}$ Fachbereich Physik, Bergische Universit\"at Gesamthochschule
          Wuppertal, Wuppertal, Germany$^a$ \\
 $ ^{34}$ DESY, Zeuthen, Germany$^a$ \\
% ...
$ ^a$ Supported by the Bundesministerium f\"ur Bildung, Wissenschaft,
        Forschung und Technologie, FRG,
        under contract numbers 7AC17P, 7AC47P, 7DO55P, 7HH17I, 7HH27P,
        7HD17P, 7HD27P, 7KI17I, 6MP17I and 7WT87P \\
% ...
$ ^f$ Supported by the Deutsche Forschungsgemeinschaft \\
% ...
$ ^k$ Supported by Russian Foundation for Basic Research 
       grant no. 96-02-00019 \\
% ...

\newpage
\section{Introduction}
\label{sec:intro}

The $ep$ collider HERA offers the unique possibility to search for 
$s$-channel production of new particles which couple to 
lepton-parton pairs. 
Examples are leptoquark (LQ) colour triplet bosons which appear naturally 
in various unifying theories beyond the Standard Model (SM) such as 
Grand Unified Theories~\cite{GUT} and Superstring inspired $E_6$ 
models~\cite{E6}, and in some Compositeness~\cite{COMPOSITE} 
and Technicolour~\cite{TECHNICO} models.
Leptoquarks could be singly produced by the fusion of the $27.5 \GeV$ 
initial state lepton with a quark of the $820 \GeV$ incoming proton, 
with masses up to the kinematic limit of 
$\sqrt{s_{ep}} \simeq 300 \GeV$.

The interest in such new bosons has been considerably renewed recently 
following the observation by the H1~\cite{H1HIQ2} and ZEUS~\cite{ZEUSHIQ2} 
experiments of a possible excess of events at very high masses and squared 
momentum transfer $Q^2$, above expectations from SM neutral current (NC) 
and charged current (CC) deep-inelastic scattering (DIS).
These early results were based on data samples collected from 1994 to 1996. 
Of particular interest was the apparent ``clustering'' of outstanding 
NC events at masses around $200 \GeV$ observed in H1 which has motivated 
considerable work on LQ kinematics~\cite{DREESBERN}, constraints and
phenomenology~\cite{HIGHXYLQ}, extending beyond the original effective 
model of Buchm\"uller-R\"uckl-Wyler (BRW)~\cite{BUCH1987}. 

In this paper, LQs are searched using all available $e^+p$ data collected 
in H1 from 1994 to 1997.
Inclusive single and double differential DIS cross-sections obtained 
from a similar dataset are presented in a separate paper~\cite{H1F2PAPER}.
Here, firstly, NC and CC measurements at high $Q^2$ are compared with SM 
expectations at detector level.
Mass and angular distributions of NC- and CC-like events 
are then used to set constraints on first generation LQs.
The search is then further extended to LQs possessing couplings
to leptons of different generations.
Such lepton flavor violating (LFV) LQs would lead to final states involving 
a second or third generation lepton.

The total integrated luminosity ${\cal{L}}$ amounts to $37 \picob^{-1}$, 
an increase in statistics of a factor $\sim 2.6$ compared to previous 
H1 analysis at very high $Q^2$~\cite{H1HIQ2} and a factor $\sim 13$
compared to previous LQ searches at HERA~\cite{H1LQ,ZEUSLQ,ZEUSLFV}.

\section{The H1 Detector}
\label{sec:detector}

A complete description of the H1 detector can be found
elsewhere~\cite{H1DETECT}.
Here we introduce only the components relevant for the present analysis
in which the final state of the processes involves either a charged
lepton\footnote{The analysis does not distinguish explicitly between
                 $+$ and $-$ charges.}
with high transverse energy or a large amount of hadronic transverse
energy flow.

Positron energies and angles are measured in a liquid argon (LAr) 
sampling calorimeter~\cite{H1LARCAL} covering the polar
angular\footnote{The $z$ axis is taken to be in the direction of the
                 incident proton, the forward direction,
                 and the origin of coordinates is
                 the nominal $ep$ interaction point.}
range 4$^{\circ} \le \theta \le$ 154$^{\circ}$ and all azimuthal angles.
The LAr calorimeter is divided in eight ``wheels" along the beam axis, 
themselves subdivided in up to 8 modules with minimum inactive material
(``cracks") in between.
The modules consist of a lead/argon electromagnetic section followed by a 
stainless steel/argon hadronic section.
Their fine read-out granularity is optimized to provide approximately 
uniform segmentation in laboratory pseudorapidity and azimuthal angle $\phi$.
Electromagnetic shower energies are measured with a resolution of
$\sigma(E)/E \simeq$ $12\%$/$\sqrt{E/\GeV} \oplus1\%$ and pion
induced hadronic energies with
$\sigma(E)/E \simeq$ $50\%$/$\sqrt{E/\GeV} \oplus2\%$ after
software energy weighting.
These energy resolutions were measured in test beams with electron 
energies up to $166 \GeV$~\cite{H1CALEPI,H1CALRES} and
pion energies up to $205 \GeV$~\cite{H1CALRES}.
The energy calibration was determined initially from test beam
data with an uncertainty of $3\%$ and $4\%$ for electromagnetic  
and hadronic energies respectively.
A new absolute energy scale calibration for positrons detected in the 
actual H1 experiment has been recently 
established~\cite{H1F2PAPER,PHDBRUEL} {\it in situ}
by using the over-constrained kinematics of NC DIS, QED Compton
and $e^+ e^-$ pair production from two-photons processes.
A precision of $0.7\%$ is reached in the LAr central barrel region 
$80^{\circ} \lsim \theta_e \lsim 145^{\circ}$, 
$1.5\%$ in $40^{\circ} \lsim \theta_e \lsim 80^{\circ}$ and 
$3.0\%$ in the forward region $5^{\circ} \lsim \theta_e \lsim 40^{\circ}$.
The precision on the hadronic energy scale was determined by requiring
the balance of the transverse momenta of the positron and hadronic
system in NC DIS events. This was performed using a method~\cite{H1F2PAPER}
correcting the energy flow associated to jets by LAr wheel
calibration constants.
The hadronic energy scale is found to be understood at the $2\%$ level 
when comparing to Monte Carlo expectation.
This represents an improved understanding of both the electromagnetic 
and hadronic energy scales compared to~\cite{H1HIQ2}, made possible 
by the increase of statistics accumulated in 1997.
All analyses described in the following rely on this updated calibration.
The resolution on the polar angle of the positron measured from the
electromagnetic shower in the calorimeter varies from $\sim 2$ mrad
below 30$^{\circ}$ to $\lsim 5$ mrad at larger angles.
A lead/scintillating-fibre backward calorimeter~\cite{H1SPACAL} extends the
coverage\footnote{The detectors in the backward region were upgraded
                  in 1995 by the replacement of the lead/scintillator 
                  calorimeter~\cite{H1BEMC} and a proportional chamber.}
at larger angles (153$^{\circ} \le \theta \lsim$ 178$^{\circ}$).

Located inside the calorimeters is the tracking system which is used here
to determine the interaction vertex and provide charged track information
relevant for lepton identification (see section~\ref{sec:kinematics}).
The main components of this system are central drift and proportional
chambers (25$^{\circ} \le \theta \le$ 155$^{\circ}$), a forward track
detector  (7$^{\circ} \le \theta \le$ 25$^{\circ}$) and a backward
drift chamber\footnotemark[3].
The tracking chambers and calorimeters are surrounded by a superconducting
solenoid providing a uniform field of $1.15${\hbox{$\;\hbox{\rm T}$}}
parallel to the $z$ axis within the detector volume.
The instrumented iron return yoke surrounding this solenoid is used to
measure leakage of hadronic showers and to recognize muons.
The luminosity is determined from the rate of the Bethe-Heitler
$e p \rightarrow e p \gamma$ bremsstrahlung measured in a luminosity
monitor. 
This consists of a positron tagger and a photon tagger located 
along the beam line, $- 33$ m and
$- 103$ m respectively from the interaction point. 

For the acquisition of events we rely on the timing information
from a time-of-flight system and on the LAr trigger system which
provides a measurement of the energy flow using coarse trigger
towers~\cite{H1LARCAL}.

\section{Leptoquark Phenomenology and Models}
\label{sec:pheno}

\begin{figure}[h]
  \begin{center}
   \epsfxsize=0.7\textwidth
   \epsffile{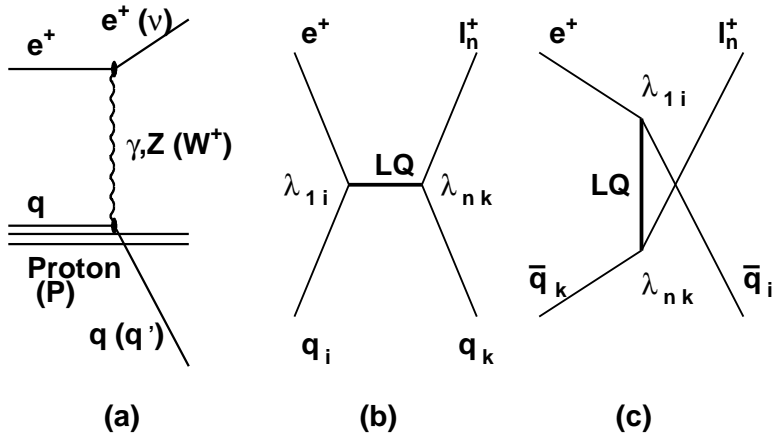}
   \caption[]{ \label{fig:diagdislq}
      {\small Diagrams of (a) deep-inelastic scattering;
              (b) $s$-channel resonant production and
              (c) $u$-channel exchange of a leptoquark with
              fermion number $F = 0$.
              Diagrams involving a $\mid F \mid = 2$ leptoquark are 
              obtained from (b) and (c) by exchanging $q$ and 
              $\bar{q}$. }}
   \end{center}
  \end{figure}

Leptoquark production at HERA can lead to final states
similar to those of DIS physics at very high $Q^2$.
The basic DIS processes are illustrated in Fig.~\ref{fig:diagdislq}a.
Leptoquarks can be resonantly produced in the $s$-channel and
exchanged in the $u$-channel as illustrated by the
diagrams in Fig.~\ref{fig:diagdislq}b,c. 
Here and in the following, whenever specified, the indices $i$ and
$j$ of the couplings $\lambda_{ij}$ at the LQ$-$lepton$-$quark 
vertices refer to the lepton $i^{th} $and quark $j^{th}$ generation 
respectively. 
Otherwise, $\lambda$ designates a coupling of LQs to first 
generation fermions.
LQs coupling only to first generation fermions (henceforth called first
generation LQs) give $e + q$ or $\nu + q'$ final states leading to 
individual events indistinguishable from SM NC and CC DIS respectively.
LQs with LFV couplings to second or third generation leptons (henceforth 
called LFV LQs) can participate in $e^+ p \rightarrow \mu^+ + jet + X$ 
or $e^+ p \rightarrow \tau^+ + jet + X$ processes.
Such exotic signatures are expected to be essentially background free 
for high transverse momentum of the observable lepton.

In the $s$-channel, a LQ is produced at a mass $M =\sqrt{s_{ep} x}$ 
where $x$ is the momentum fraction of the proton carried by the struck quark.
Over a large fraction of the mass range accessible at HERA 
and for a reasonable coupling strength, e.g. satisfying 
$\lambda^2 / 4 \pi < 1$, the intrinsic decay width of a scalar (S) or vector (V) LQ 
of nominal mass $M_{LQ}$ into a lepton and a quark is expected to be small. 
This width is calculated as
     $\Gamma_S = (3/2) \Gamma_V = \lambda^2_{ij} M_{LQ} / 16 \pi$
which corresponds for example to $\Gamma_S \simeq 40 \MeV$ for a scalar
at $M_{LQ} = 200 \GeV$ and $\lambda_{ij} = 0.1$.
In the narrow-width approximation (NWA), the resonant production 
cross-section $\sigma_{\rm{NWA}}$ is proportional to $\lambda^2 q(x)$ where  
$q(x)$ is the density of the struck parton in the incoming proton.
However when approaching the kinematic limit where the values of $q(x)$ 
are very small, the coupling strengths which can be probed with the 
actual integrated luminosities are too high for the NWA to be valid.
The convolution of the steeply falling $q(x)$ with the Breit-Wigner 
distribution of finite width characterizing the resonance leads 
to a strong distortion of the LQ mass peak, and the mass spectrum shows very 
large tails towards low values.
As a result the LQ production cross-section $\sigma_s$
in the $s$-channel for $M_{LQ}$ approaching $\sqrt{s_{ep}}$ is
considerably larger than $\sigma_{\rm{NWA}}$. 
The deviation from $\sigma_{\rm{NWA}}$ is significant (typically $> 10\%$) at
$M_{LQ} \, \gsim \, 250 \GeV$ for a LQ produced via a valence quark 
($u,d$), and already at $M_{LQ} > 200 \GeV$ for a production via a sea
quark ($\bar{u}, \bar{d}$)~\cite{DURHAM}.
The analysis presented in the following fully takes into account
these effects originating from the finite LQ decay width.

Scalar LQs produced in the $s$-channel decay isotropically in their rest 
frame leading to a flat ${\rm d} \sigma\,/\,{\rm d}y \;$ spectrum where
$y= Q^2 / s_{ep}x = \frac{1}{2}\left(1+\cos{\theta^*}\right)$ is the 
Bjorken scattering variable in DIS and $\theta^*$ is the decay polar 
angle of the lepton relative to the incident proton in the LQ centre of 
mass frame.
In contrast, events resulting from the production and decay of
vector LQs would be distributed according to 
${\rm d} \sigma\,/\, {\rm d} y \propto \, (1-y)^2$.
These $y$ spectra (or in other words the specific angular distributions 
of the decay products) from scalar or vector LQ production are markedly 
different from the ${\rm d} \sigma\,/\, {\rm d} y \propto \,y^{-2}$
distribution expected at fixed $x$ for the dominant $t$-channel
photon exchange in neutral current DIS events
\footnote{At high momentum transfer, $Z^0$ exchange is no longer
          negligible and contributes to less pronounced differences
          in the $y$ spectra between LQ signal and DIS background.}.
Hence, a LQ signal in the NC-like channel will be statistically most prominent
at high $y$.

The $u$-channel contribution scales with $\lambda^4$.
It can compete with resonant production only for LQs with fermion number 
$\mid F \mid = 2$ and at high couplings and LQ masses.
For $F=0$ LQs, it is highly suppressed by less favorable parton densities
as it proceeds via an exchange involving an {\it{antiquark}} from the 
proton.
Scalar LQ exchange would lead to events distributed in $y$ according 
to ${\rm d} \sigma\,/\, {\rm d} y\sim\, (1-y)^2$ while vector LQ 
exchange would lead to a flat $y$ spectrum.
However the events originating
from $u$-channel LQ exchange would mainly be concentrated at
mass values much lower than $M_{LQ}$.
As such, the kinematic cuts used in this analysis to reduce the
number of NC-like and CC-like events (see section~\ref{sec:kinematics})
also drastically suppress a possible $u$-channel contribution.

In approaching $M_{LQ} \sim \sqrt{s_{ep}}$, the interference of the
LQ $s$-channel production and $u$-channel exchange with
SM boson exchange can no longer be
neglected. 
This interference can be constructive or destructive\footnote{
    The signs of the interference terms between SM gauge boson and 
    $F=0$ LQ contributions given in the original BRW paper~\cite{BUCH1987} 
    were found to be incorrect. The correct signs as provided in the
    {\it erratum} to Ref.~\cite{BUCH1987} have been used here. }
depending on the LQ type.
As will be seen in section~\ref{sec:lqfirst}, the set of cuts used 
in the present analysis focuses on a phase space region where the 
contribution of the interference is considerably reduced.

\begin{table*}[htb]
  \renewcommand{\doublerulesep}{0.4pt}
  \renewcommand{\arraystretch}{1.2}
  \vspace{-0.1cm}
  
  \begin{center}
  \begin{tabular}{c|c|c|c|c|}
    \cline{2-5}
     & \multicolumn{2}{|c|}{Angular Spectrum} 
                                 & \multicolumn{2}{c|}{Cross-Section} \\
     & \multicolumn{2}{|c|}{$y$ Shape} 
                                 & \multicolumn{2}{c|}{$\lambda$ Dependence} \\
    \cline{2-5}
     &  Scalar      & Vector           
                            & $M \ll \sqrt{s_{ep}}$ & $ M \gg \sqrt{s_{ep}}$ \\
    \hline
  \multicolumn{1}{|c|}{$s$-channel}
     &  flat        & $(1-y)^2$           & $\lambda^2_{1i} \beta_n$     
                                    & $\lambda^2_{1i} \lambda^2_{nj}$     \\
    \hline
  \multicolumn{1}{|c|}{$u$-channel}
     & $(1-y)^2$    & flat                & $\lambda^2_{1i} \lambda^2_{nj}$ 
                                    & $\lambda^2_{1i} \lambda^2_{nj}$     \\
     \hline
  \multicolumn{1}{|c|}{Interference}
     & \multicolumn{2}{c|}{}        & $\lambda^2$          & $\lambda^2$     \\
    \hline
  \end{tabular}
  \caption {\small \label{tab:su_properties}
           Main properties of the different LQ induced contributions at HERA
           to $e^+ +  q_i \rightarrow l^+_n + q_j$.
           For LQs coupling to both $e q$ and $l_n q$ pairs
           with $l_n \ne e$, $\beta_n$
           denotes the branching ratio of the LQ into $l_n + q_j$.
           The interference contribution only concerns 
            processes with a first generation lepton in the final state. }
  \end{center}
\end{table*}

The experiments at HERA are also sensitive to LQs with 
$M_{LQ} \, \gsim \, \sqrt{s_{ep}}$.
For first generation LQs, the interference between LQ induced 
and SM boson exchange processes (which scales with $\lambda^2$) generally 
dominates in this mass range over the $s$-and $u$-channel contributions 
(both scaling with $\lambda^4$).
For $F=0$ LQs, a nevertheless sizeable $s$-channel contribution originates 
from the convolution of the parton density with the low mass tail of the
LQ Breit-Wigner resonance of finite width, provided that the
coupling $\lambda$ is not too small. 
For example, for $\lambda =1$ and for the LQ labelled $S_{1/2,L}$
in the BRW model (see below),
the $s$-channel contribution competes with the interference
for $M_{LQ}$ up to $\simeq 400 \GeV$, within the kinematic cuts
used in section~\ref{sec:brwlim}.

In the cases of $M_{LQ} \gg \sqrt{s_{ep}}$, the propagator entering the
LQ amplitudes can be contracted to a four-fermion interaction.
One is left with a contact interaction mostly affecting the measured
inclusive DIS $Q^2$ spectrum through interference effects.
Constraints on such four-fermion couplings translated into limits
on $M_{LQ}/\lambda$ for first generation LQs will be discussed in a separate
paper.
For LFV LQs above $\sqrt{s_{ep}}$, both the $s$- and $u$-channel 
contributions may in principle be important for large Yukawa coupling 
values.
There, the LFV LQ cross-sections
$\sigma(e q_i \rightarrow l_n q_j)$ ($s$-channel)
and $\sigma(e \bar{q}_j \rightarrow l_n \bar{q}_i)$ ($u$-channel)
only depend on $\lambda_{1i}$, $\lambda_{nj}$ and $M_{LQ}$ via 
$\lambda_{1i}^2 \lambda_{nj}^2 / M^4_{LQ}$.

Some essential characteristics of the different LQ induced processes
contributing at HERA are summarized in table~\ref{tab:su_properties}.

The LQ searches will be discussed here either in the strict context of 
the BRW phenomenological ansatz~\cite{BUCH1987} where the decay 
branching ratios are fixed by the model, or in the context of generic 
models allowing for arbitrary branching ratios.  
The BRW model considers all possible scalar ($S_I$) and vector ($V_I$) 
LQs of weak isospin {\it I} with dimensionless couplings $\lambda^{L,R}_{ij}$ 
to lepton-quark pairs, where
{\it L} or {\it R} is the chirality of the lepton.
The general effective Lagrangian which is introduced obeys the 
symmetries of the SM.
There are 10 different LQ isospin multiplets, with couplings
to left or right handed fermions, among which there are 5 isospin
families of scalar LQs.
These are listed in table~\ref{tab:brwscalar}.
\begin{table*}[htb]
  \renewcommand{\doublerulesep}{0.4pt}
  \renewcommand{\arraystretch}{1.2}
 \vspace{-0.1cm}

\begin{center}
    \begin{tabular}{|c|c|c||c|c|c|}
      \hline
       $F=-2$ & Prod./Decay & $\beta_e$
              & $F=0$ & Prod./Decay & $\beta_e$  \\

      \hline
%
     \multicolumn{6}{|c|}{Scalar Leptoquarks} \\ \hline
    $^{1/3}S_0$     & $e^+_R \bar{u}_R\rightarrow e^+ \bar{u}$ & $1/2$
  & $^{5/3}S_{1/2}$ & $e^+_R u_R \rightarrow e^+ u$            & $1$  \\
                          & $e^+_L \bar{u}_L\rightarrow e^+ \bar{u}$ & $1$
  &                       & $e^+_L u_L \rightarrow e^+ u$            & $1$ \\
      \cline{1-3}
      $^{4/3}\tilde{S}_0$
        & $e^+_L \bar{d}_L\rightarrow e^+ \bar{d}$ & $1$
  & $^{2/3}S_{1/2}$ & $e^+_L d_L \rightarrow e^+ d$            & $1$ \\
      \hline
      $^{4/3}S_1$
        & $e^+_R \bar{d}_R \rightarrow e^+ \bar{d}$
         & $1$
  & $^{2/3}\tilde{S}_{1/2}$ & $e^+_R d_R \rightarrow e^+ d$ & $1$ \\
      $^{1/3}S_1$
        & $e^+_R \bar{u}_R \rightarrow e^+ \bar{u}$
         & $1/2$
             & & &  \\
      \hline
%
     \multicolumn{6}{|c|}{Vector Leptoquarks} \\ \hline
    $^{4/3}V_{1/2}$ & $e^+_L \bar{d}_R\rightarrow e^+  \bar{d}$ & $1$  
  & $^{2/3}V_{0}$   & $e^+_L d_R \rightarrow e^+ d$              & $1$ \\
                          & $e^+_R \bar{d}_L\rightarrow e^+  \bar{d}$ & $1$
  &                       & $e^+_R d_L \rightarrow e^+ d$             & $1/2$ \\
      \cline{4-6}
    $^{1/3}V_{1/2}$ & $e^+_L \bar{u}_R\rightarrow e^+  \bar{u}$ & $1$ 
  & $^{5/3}\tilde{V}_0$
        & $e^+_L u_R \rightarrow e^+ u$ & $1$ \\
      \hline
    $^{1/3}\tilde{V}_{1/2}$
        & $e^+_R \bar{u}_L\rightarrow e^+ \bar{u}$ & $1$ 
  & $^{5/3}V_{1}$    & $e^+_R u_L \rightarrow e^+ u$              & $1$ \\
                          &                                            &
  & $^{2/3}V_{1}$    & $e^+_R d_L \rightarrow e^+ d$              & $1/2$ \\
      \hline
      \hline
    \end{tabular}
    \caption {\small \label{tab:brwscalar}
               Leptoquark isospin families in the Buchm\"uller-R\"uckl-Wyler 
               model.
               For each leptoquark, the superscript corresponds to its 
               electric charge, while the subscript denotes its weak 
               isospin.
               For simplicity, the leptoquarks are conventionally indexed 
               with the chirality of the incoming {\it{electron}} which 
               could allow their production in $e^-p$ collisions, 
               e.g. the $\tilde{S}_0$ will be denoted by $\tilde{S}_{0,R}$
               (see text). $\beta_e$ denotes the branching ratio of the
               LQ into $e^+ + q$. }

\end{center}
\end{table*}
We restrict the search to pure chiral couplings of the LQs given that 
deviations from lepton universality in helicity suppressed pseudoscalar 
meson decays have not been observed~\cite{DAVIDSON,LEURER}. 
This restriction to couplings with either left- ($\lambda^L$) or
right-handed ($\lambda^R$) leptons 
(i.e. $\lambda^L \cdot \lambda^R \sim 0$),
affects only two scalar ($S_0$ and $S_{1/2}$) and two vector
($V_{1/2}$ and $V_0$) LQs.
We make use of the so-called Aachen nomenclature and classification
scheme~\cite{HERAWS}
and do not use specific symbols to label the {\it{anti-leptoquarks}}
which are actually produced in $e^+ p$ collisions.
We make the simplifying assumptions that one of the LQ multiplets
is produced dominantly and that the mass eigenstates within the
LQ isospin doublets and triplets are degenerate in mass.

For the determination of LQ signal detection efficiencies,
we make use of the LEGO event generator~\cite{LEGO}
and of a complete simulation of the H1 detector response.
LEGO incorporates $s$- and $u$-channel LQ exchange processes depicted in 
Fig.~\ref{fig:diagdislq}b,c.
It takes into account initial state QED radiation in the collinear 
approximation.
The parton showers approach~\cite{JETSET74} relying 
on the DGLAP~\cite{DGLAP} evolution 
equations is used
to simulate QCD corrections in the initial and final states,
and the kinematics at the decay vertex is properly corrected
for effects of the parton shower masses.
The non-perturbative part of the hadronization is simulated 
using string fragmentation~\cite{JETSET74}.
The Mandelstam variable $\hat{s}$ characterizing the $eq \rightarrow lq$ 
subprocess defines the scale at which the parton density is evaluated 
as well as the maximum virtuality of parton showers.
The LQ signal cross-sections are calculated using
the full matrix elements given in Ref.~\cite{BUCH1987} and taking into
account the contributions from the $s$- and $u$-channels as well
as the interference with SM boson exchange.
The resulting cross-sections are further corrected to account for 
next-to-leading order (NLO) QCD effects making use of multiplicative
$K$-factors~\cite{SPIRA} in a procedure described in detail in
section~\ref{sec:brwlim}. These NLO QCD corrections which depend on the
LQ signal shape expected for a given $M_{LQ}$ and $\lambda$ are typically
of ${\cal{O}}(10 \%)$.
For the parton densities, use is made of the recent 
Martin-Roberts-Stirling-Thorne MRST~\cite{MRSTSF}
parametrization which better describes existing measurements constraining 
the sea quark densities in the proton~\cite{NA51,DRELLYAN}.

The theoretical uncertainty on the signal cross-section originating
mainly from contributions of parton density distributions
extracted from ``QCD fits'' and the value of the strong coupling
constant $\alpha_S$ is treated as a systematic error.
This uncertainty is $\simeq 7 \%$ for leptoquarks coupling
to $e^+ u$, and varies between $\simeq 7 \%$ at low LQ masses up
to $\simeq 30 \%$ around $250 \GeV$ for leptoquarks
coupling to $e^+ d$. Above $250 \GeV$, for coupling values 
corresponding to the expected sensitivity, the small
but finite width of the resonance results in the fact that
mainly relatively low $x$ partons are involved in the LQ production.
Hence the uncertainty on the signal cross-section decreases
for $F=0$ LQs to $\simeq 7 \%$ between $250 \GeV$
and the kinematic limit.
For $\mid F \mid =2$ leptoquarks, this uncertainty ranges
from $\simeq 10 \%$ at low masses and reaches
$\simeq 40 \%$ around $200 \GeV$, and then goes down
to $\simeq 15 \%$ at the kinematic limit.
Moreover, choosing alternatively $Q^2$ or the square of the transverse 
momentum of the final state lepton instead of $\hat{s}$ as the hard 
scale at which the parton distributions are estimated yields an
additional uncertainty of $\pm 7 \%$ on the signal cross-section.

\section{Deep Inelastic Scattering and Other Background Sources}
\label{sec:dismc}
 
The calculation of the SM expectation for NC and CC DIS $ep$ 
scattering is performed using the parton model in the approximation of 
single $\gamma/Z$ and $W$ boson exchange, and relies on a description of 
the proton in terms of scale dependent structure functions. 
The structure functions are expressed in terms of parton densities and
are taken here from the MRST parametrization 
which includes constraints from HERA data up to
$Q^2 = 5000 \GeV^2$~\cite{H1MRST,ZEUSMRST}.
The parton densities are evolved to the high $Q^2$ domain relevant for 
this analysis using the next-to-leading order DGLAP equations.
The Monte Carlo event generator DJANGO~\cite{DJANGO} which follows such 
an approach is used for the comparison with data.
This generator includes the QED first order radiative
corrections~\cite{HERACLES} and a
modelling of the emission of QCD radiation via ARIADNE~\cite{ARIADNE}.
The ARIADNE generator makes use of the Colour Dipole
Model~\cite{CDM} to simulate QCD radiation to all orders and 
string fragmentation to generate the hadronic final 
state. 

The contributions from all background processes which could give
rise to events with true or misidentified isolated leptons at high 
transverse energy or to events with a large missing transverse momentum 
have been evaluated.
In particular, direct and resolved photoproduction processes were modelled 
using the PYTHIA generator~\cite{PYTHIA}.
It is based on leading order QCD matrix elements and includes initial 
and final state parton showers calculated in the leading logarithm 
approximation, and string fragmentation.
The renormalization and factorization scales were both set to $P_T^2$,
$P_T$ being the transverse momentum of the jets emerging out of the
hard subprocess.
The GRV~(G) leading order parton densities in the 
proton (photon) have been used~\cite{GRVG}.
The production of electroweak vector bosons $Z^0$ and
$W^{\pm}$ was modelled using the EPVEC~\cite{EPVEC} event generator.
Contributions from two-photon processes where one $\gamma$ originates
from the proton were also considered and estimated using the
LPAIR~\cite{LPAIR} event generator.
A complete Monte Carlo simulation of the H1 detector response
has been performed for all background processes.

The following experimental errors are propagated as
systematic errors on the mean SM expectations~:
\begin{itemize}
 \item the uncertainty on the integrated luminosity ($\pm 1.5 \%$);
  \item the uncertainty on the absolute calibration of the calorimeters
        for electromagnetic energies, ranging between $\pm 0.7\%$ in the 
        central LAr wheels to $\pm 3\%$ in the forward region 
        of the LAr calorimeter (see section~\ref{sec:detector});
  \item the uncertainty on the calibration of the calorimeters
        for hadronic showers of $\pm 2\%$ 
        (see section~\ref{sec:detector}).
\end{itemize}

In addition, a $7\%$ theoretical uncertainty on the predicted
NC DIS cross-section originates mainly from the lack of knowledge on the
proton structure (see detailed discussion in~\cite{H1HIQ2})
and, to a lesser extent, from the higher order QED corrections.
For CC DIS processes which are mainly induced by $d$ quarks, this
uncertainty varies with $Q^2$ and ranges between $7 \%$ and
$\simeq 20 \%$ at the highest $Q^2$ considered here.
All analyses described in the following sections have been
repeated with an independent shift of the central values by $\pm 1$
standard deviation of each of the experimental and theoretical sources of
errors. The overall systematic error of the SM prediction
is determined as the quadratic sum of the resulting errors
and of the statistical error on the Monte Carlo simulation.

\section{Event Selection and Comparison with Standard Model Expectation}
\label{sec:kinematics}

The search for first generation LQs relies essentially on an
inclusive NC selection requiring an identified positron at high transverse 
energy, and an inclusive CC selection requiring a large missing transverse 
momentum. The selection of NC- and CC-like events will be described 
in subsections~\ref{sec:selectnc} and~\ref{sec:selectcc} respectively where 
the resulting samples will be compared to SM expectations.
For LQs possessing couplings to mixed fermion generations,
the selection of $e^+ p \rightarrow \mu^+ + q + X$ and
$e^+ p \rightarrow \tau^+ + q + X$ candidates which will be discussed in 
subsections~\ref{sec:selectmu} and~\ref{sec:selectau} respectively,
requires essentially an identified $\mu$ or $\tau$ lepton together with a 
large amount of hadronic transverse energy.

In common for all channels analysed, the events must have been
accepted by a LAr trigger asking either for an electromagnetic cluster,
for a large transverse energy in the central part of the calorimeter,
or for a large imbalance in the transverse energy flow.
The rejection of background from cosmic rays and from ``halo'' muons 
associated with the proton beam mainly relies on constraints on the event 
timing relative to the nominal time of the beam bunch crossings.
Beam-wall and beam-residual gas interactions are furthermore suppressed
by requiring a primary vertex in the range $\mid z - \bar{z} \mid < 40 \cm$ 
where $\bar{z}$ varies within $\pm 5 \cm$ around $z=0$ depending on 
the HERA beam settings. 

In what follows, unless explicitly stated otherwise, the energy flow 
summations run over all energy deposits $i$ in the calorimeters 
(apart from the electron and photon taggers).
Thus, the missing transverse momentum $P_{T,miss}$ is obtained as 
     $P_{T,miss} \equiv  \sqrt{ \left(\sum E_{x,i} \right)^2
                      +  \left(\sum E_{y,i} \right)^2 } $
with $ E_{x,i} = E_i \sin \theta_i \cos \phi_i $
and  $ E_{y,i} = E_i \sin \theta_i \sin \phi_i $.
The momentum balance with respect to the incident positron is
obtained as 
     $\sum \left(E - P_z\right) \equiv  \sum \left(E_i - E_{z,i}\right) $ 
with $ E_{z,i} = E_i \cos \theta_i$.

\subsection{Neutral Current Deep-Inelastic-like Signatures}
\label{sec:selectnc}

\subsubsection{Event selection and kinematics}
\label{sec:selectnckine}

The {\bf selection of NC DIS-like events} uses mainly calorimetric 
information for electron finding and energy-momentum conservation 
requirements, with selection cuts similar to those considered in
previous high $Q^2$ analysis~\cite{H1HIQ2}:
\begin{enumerate}
  \item an isolated positron with $E_{T,e} > 15 \GeV$ 
        ($E_{T,e} = E_e \sin \theta_e$),
        found within the polar angular range 
        $5^{\circ} \le \theta_e \le  145^{\circ}$;
        the positron energy cluster should contain more than $98\%$ of 
        the LAr energy found within a pseudorapidity-azimuthal cone 
        of opening $\sqrt{ (\Delta \eta_e)^2 + (\Delta \phi_e)^2 } = 0.25$
        where $\eta_{e} = -\ln \tan \frac{\theta_e}{2}$; 
        at least one charged track is required within the positron
        isolation cone;
  \item a total transverse momentum balance 
        $P_{T,miss}/\sqrt{E_{T,e}} \leq 4 \sqrt{\GeV}$ ;
  \item a limited reconstructed momentum loss in the direction of the incident
          positron such that  
         $ 40 \GeV \leq \sum \left(E - P_z\right) \leq 70 \GeV$.
\end{enumerate}
The identification of positron induced showers relies on the detailed knowledge 
of the expected lateral and longitudinal shower 
properties~\cite{H1CALEPI,PHDBRUEL}.
The efficiency for the detection of positrons exceeds $90\%$ everywhere 
within the acceptance cuts,   
the main losses being due to showers developing through the inactive 
material between calorimeter modules.
The cut (2) makes possible a very efficient NC DIS selection up to the 
highest $Q^2$ by taking into account the natural dependence of the 
calorimetric energy resolution, $E_{T,e}$ being used as an estimate
of the scale relevant for the actual $P_{T,miss}$ measurement.
The cut (3) retains more than $90\%$ of NC DIS events and exploits 
the fact that by energy-momentum conservation, the 
$\sum \left(E - P_z\right)$ distribution for NC DIS events is peaked at 
$2 E_e^0$,
where $E^0_e$ is the positron beam energy. 
It rejects events where a very hard collinear $\gamma$ is emitted 
by the initial state positron.
To ensure a good control of the positron identification performances,
a fiducial cut is applied requiring:
\begin{enumerate}
\setcounter{enumi}{3}
  \item 
        an azimuthal impact of the track associated to the positron at 
        $\mid \phi_e - \phi_{crack} \mid > 1^{\circ}$ 
        from the nearest projective $\phi$ crack in the transverse plane.
\end{enumerate}  

The DIS Lorentz invariants $Q^2, y$ and $M$ are
determined using only the measurement of the ``scattered'' positron
energy and angle as soon as $\mid \phi_e - \phi_{crack} \mid > 2^{\circ}$,
such that the measurement of $E_e$ is reliable~:
  $$ M_e = \sqrt{\frac{Q^2_e}{y_e}}, \;\;\;\;
      Q^2_e = \frac{E^2_{T,e}}{1-y_e}, \;\;\;\;
       y_e = 1 - \frac{E_e - E_e \cos \theta_e}{2E_e^0} \;\; . \;\; $$
This method will henceforth be called the electron method ($e$-method).
In  $\sim 4 \%$ of the acceptance corresponding to the range
$ 1^{\circ} < \mid \phi_e - \phi_{crack} \mid < 2^{\circ}$
where calorimetry measurements of positrons deteriorate,
the reconstructed positron energy is corrected to the value 
given by the double angle method~\cite{HOEGER}~:
$$ E_{2 \alpha} = \frac{ 2 E^0_e} {\alpha_e + \alpha_h}
                  \frac{1} {\sin \theta_e} \qquad ,$$
using 
$$\alpha_e = \tan (\theta_e/2) = \frac{E_e - E_{z,e}} {E_{T,e}} 
  \qquad {\mbox{and}} \qquad
  \alpha_h = \tan (\theta_h/2) = \frac{\sum_h (E - P_z) }
                {\sqrt{ (\sum_h E_{x})^2 + (\sum_h E_{y})^2 }} \quad , $$
where the summations run over all energy deposits of the hadronic
final state.
 
In the following analysis, the comparison with SM expectation is 
restricted to the kinematic range $Q^2 > 2500 \GeV^2$ and $0.1 < y < 0.9$.
The resolution in $M_e$ degrades with decreasing $y_e$ 
($ \delta M_e / M_e \propto 1/ y_e$) and so the low 
$y$ domain is excluded.
Excluding the high $y$ values avoids the region where migrations effects due 
to QED radiation in the initial state are largest for the $e$-method.
In the kinematic range considered and given cuts (1) to (4), the NC trigger 
efficiency exceeds $98\%$ and is consistent with $100\%$ to within
experimental error.

The $y < 0.9$ restriction also suppresses the photoproduction background
where e.g. a jet has been misidentified as an electron.
Following~\cite{H1HIQ2}, any possibly remaining non-DIS contamination 
coming for example from $\gamma\gamma$ or QED Compton processes as well 
as the background from misidentified low $Q^2$ NC DIS are further 
suppressed through a minimal set of specific cuts~\cite{PHDBRUEL}.
Among these a prominent one against multi-lepton final states is the
requirement of at least one reconstructed jet with $E_{T,jet} > 7 \GeV$
found using a cone algorithm in the laboratory frame, with a radius
$ \sqrt{ (\Delta \eta)^2 + (\Delta \phi)^2 } = 1$.
The jet should be in the polar angular range 
$7^{\circ} \le \theta_{jet} \le  145^{\circ}$, and
at least 5\% of its energy should be deposited in the hadronic
section of the LAr calorimeter.
Against photoproduction and low $Q^2$ NC DIS, it is also required 
that there be less than $7.5 \GeV$ in the backward calorimeter.
The specific reduction of the background contamination induces only 
small ($< 5\%$) efficiency loss for high $Q^2$ NC DIS-like processes. 
The remaining contamination is estimated to be below $0.15\%$ and is 
henceforth neglected.

Leptoquark signal selection efficiencies are determined over a
coupling-mass grid with steps in mass of $25 \GeV$ and for
coupling values corresponding roughly to the expected 
sensitivity to properly take into account the effect
of the intrinsic finite width of the searched resonance.
Detailed Monte Carlo simulation of about 500 events per point on
the grid is performed followed by the application of the full
analysis chain.
The above set of cuts ensures a typical selection efficiency
for $LQ \rightarrow e+q$ events which varies between
$\simeq 40 \%$ and $\simeq 75 \%$ for LQ masses ranging
in $75$ to $250 \GeV$.

Applying all the above NC selection criteria, $1298$ DIS event
candidates are accepted which is in good agreement with the expectation of 
$1243 \pm 95$ events from standard NC DIS.

\subsubsection{Comparison with Standard Model expectation}
\label{sec:DISNCsample}

\begin{figure}[htb]
   \begin{center}
     \epsfxsize=1.0\textwidth
     \epsffile{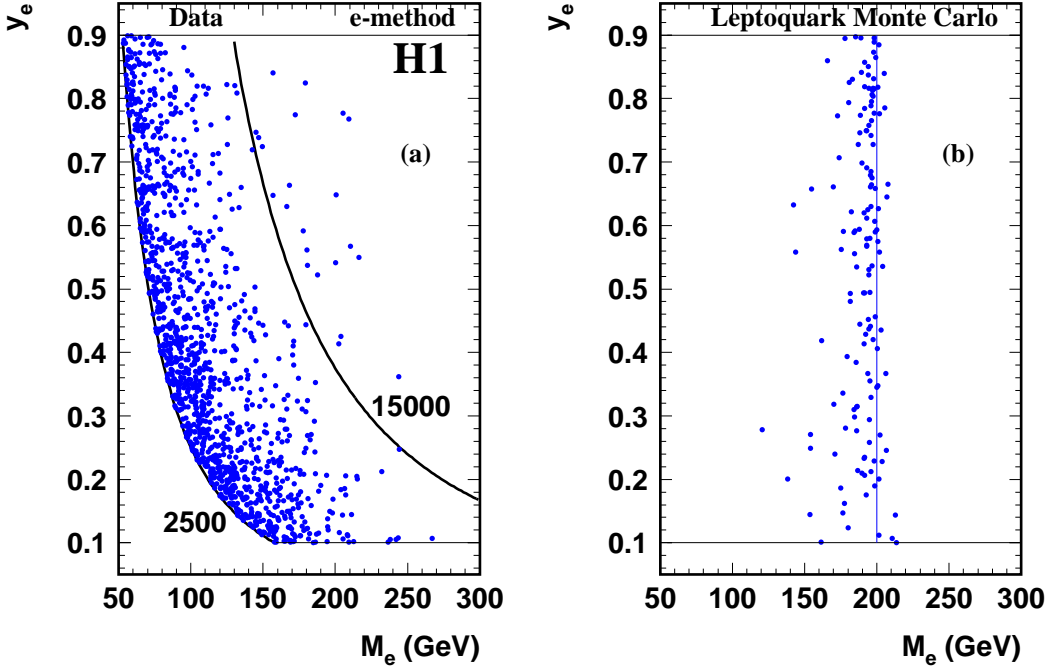}
      \caption
         {\small \label{fig:scatter}
                 Kinematics in the $y_e$ - $M_e$ plane of
                 (a) the selected NC DIS candidate events from H1 data
                     (two isocurves at $Q^2 = 2500$ and $15000 \GeV^2$
                     are plotted as full lines);
                 (b) a scalar $F=0$ leptoquark of mass 
                     $M_{LQ} = 200 \GeV$ decaying into $e+q$, for
                     a coupling $\lambda = 0.05$.}
 \end{center}
\end{figure}

Figure~\ref{fig:scatter}a shows the distribution of the NC candidates in
the $y_e$ - $M_e$ kinematic plane.
In such a plane, the signal of a $200 \GeV$ scalar LQ with
$F=0$ could manifest as illustrated in Fig.~\ref{fig:scatter}b,
for coupling values corresponding to the expected sensitivity.
Compared to other commonly used kinematic methods for NC DIS
at HERA~\cite{H1F2PAPER}, the $e$-method provides the best peak resolution 
(truncated Gaussian fit) in mass at high $y$, where a LQ signal would be most 
prominent.
This resolution on $M_e$ varies within $\simeq 3 - 6 \GeV$ for LQ masses 
ranging between 100 and $250 \GeV$.
It should be noted, as can be inferred from Fig.~\ref{fig:scatter}b,
that the $e$-method underestimates on average the true LQ mass by 
$\simeq 2 \%$ due to migrations caused by final state QCD radiation.
%
%

A differential analysis in the $y_e$-$M_e$ plane of the very high $Q^2$
events from the 1994 to 1996 datasets~\cite{H1HIQ2} had revealed
a noteworthy excess of NC DIS-like events at $Q^2_e \, \gsim \, 15000 \GeV^2$
or for $y_e > 0.4$ at masses $M_e \simeq 200 \GeV$.
A comparison of the measured $M$, $y$ and $Q^2$ spectra with standard DIS 
model expectations can now be re-examined with higher statistics.

\begin{figure}[htb]
 \begin{center}
   \epsfxsize=0.8\textwidth
   \epsffile{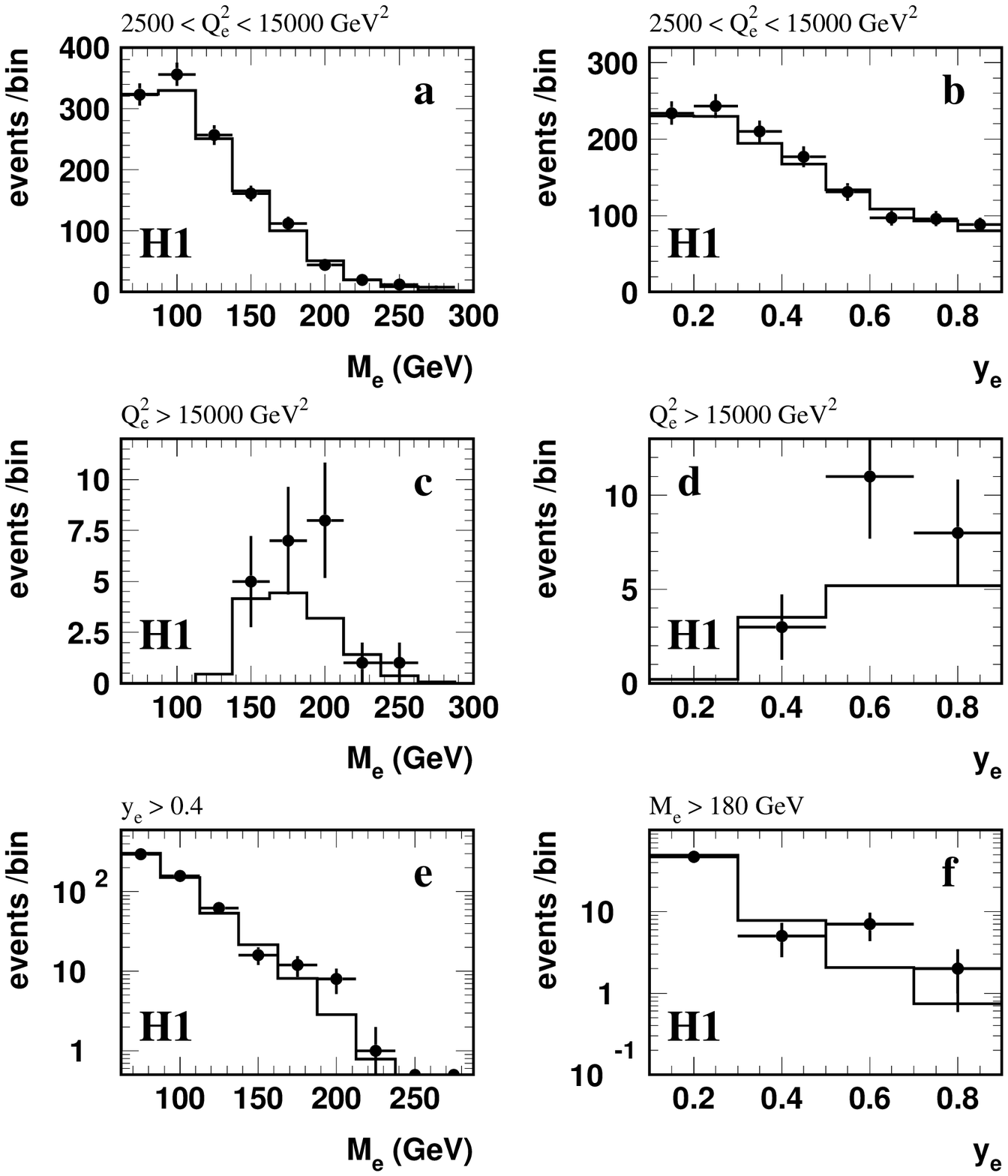}
 \caption[]{ \label{fig:mandy}
    { \small Distributions of $M_e$ and $y_e$ for the selected NC DIS 
             candidate events, 
             (a) and (b) for $2500 < Q^2_e < 15000 \GeV^2$,
             (c) and (d) for $Q^2_e > 15000 \GeV^2$;
             distributions of (e) $M_e$ for $y_e > 0.4$ and (f) $y_e$ for 
             $ M_e > 180 \GeV $; superimposed on the data points 
             ($\bullet$ symbols) are histograms of the standard NC DIS
             expectation. }}
 \end{center}
\end{figure}

We first consider the $M_e$ and $y_e$ information.
The projected $M_e$ and $y_e$ spectra are shown in Fig.~\ref{fig:mandy} in 
several kinematic domains.
Figures~\ref{fig:mandy}a and~\ref{fig:mandy}b show the projected $M_e$ and
$y_e$ distributions of the NC DIS-like selected events at ``moderate'' $Q^2$
($2500 < Q^2_e < 15000 \GeV^2$) and Fig.~\ref{fig:mandy}c 
and~\ref{fig:mandy}d at ``very high'' $Q^2$ ($Q^2_e > 15000 \GeV^2$).
The distributions of the measured data are well reproduced by standard DIS
predictions in the low $Q^2$ range.
%
%
At high $Q^2$ the data slightly exceed the NC DIS expectation
at $M_e \sim 200 \GeV$ as can be seen in Fig.~\ref{fig:mandy}c.
Moreover, Fig.~\ref{fig:mandy}d shows that the excess of observed
events is more prominent at high $y_e$, so that
at high $M_e$ and large $y_e$ the experiment tends to exceed the SM
expectation.

Figure~\ref{fig:mandy}e shows the measured and expected $M_e$
distributions for a minimum $y_e$ value of $y_{min} = 0.4$.
An excess of events over the NC DIS expectation at high mass
($\sim 200 \GeV$) is still visible.
In the mass range $ 200 \GeV \pm \Delta M / 2$ with $\Delta M = 25 \GeV$,
$N_{obs} = 8$ events are observed for an expectation of
$N_{DIS} = 2.87 \pm 0.48$.
The mean mass value of these 8 events as determined
with the $e$-method $ \langle M_e \rangle = 202.5 \pm 7.0 \GeV$
(RMS) agrees within $1.4 \%$ with the one obtained
from the invariant mass of the final $e$-jet pairs.
Of the observed events, 5 originate from the 1994 to 1996 data
($40.3\%$ of ${\cal{L}}$) and 3 from the 1997 data ($59.7\%$ of ${\cal{L}}$).
It should be emphasized here that $N_{obs}$ and $N_{DIS}$ are
quoted for the same $\Delta M - \Delta y$ region where the most significant
excess was observed in the original analysis of the
1994 to 1996 data~\cite{H1HIQ2} despite the fact that the
individual events are slightly (within originally estimated systematic
errors) displaced in the $M-y$ plane.
In this domain, 7 events were reported in~\cite{H1HIQ2}.
These events are measured here at $M_e$ values on average $2.4\%$ higher
due to the new {\it in situ} calibration of the electromagnetic section
of the LAr calorimeter, and thus, one event has now migrated
outside this $\Delta M$ domain.
The estimated mass of one of the other 6 events in which the positron lies
within less than $2^{\circ}$ from the closest $\phi$ crack
was and remains measured outside this mass region when using the
double angle method in contrast to the $e$-method used in~\cite{H1HIQ2}.
It was explicitly checked that repeating the 1994 to 1996 
analysis procedures of~\cite{H1HIQ2} but using this
new calibration leaves the statistical significance
and the physics messages of~\cite{H1HIQ2} unchanged.

At large mass $M_e > 180 \GeV$ and for $y_e > 0.4$, we observe
in the 1994 to 1996 data $N_{obs} = 7$ in slight excess of
the expectation of $N_{DIS} = 2.21 \pm 0.33$ while in the 1997 data
alone $N_{obs} = 4$ events are observed, in good agreement with
the expectation of $N_{DIS} = 3.27 \pm 0.49$.
The $y_e$ distribution of these high mass events is shown
in Fig.~\ref{fig:mandy}f.

Hence, no significant excess is seen in the mass spectrum for the
1997 data alone and the ``clustering'' around $M_e \sim 200 \GeV$
is, overall, thus rendered less significant than that observed with
1994 to 1996 data only.

\begin{figure}[htb]
  \begin{center}
    \epsfxsize=1.0\textwidth
  \epsffile{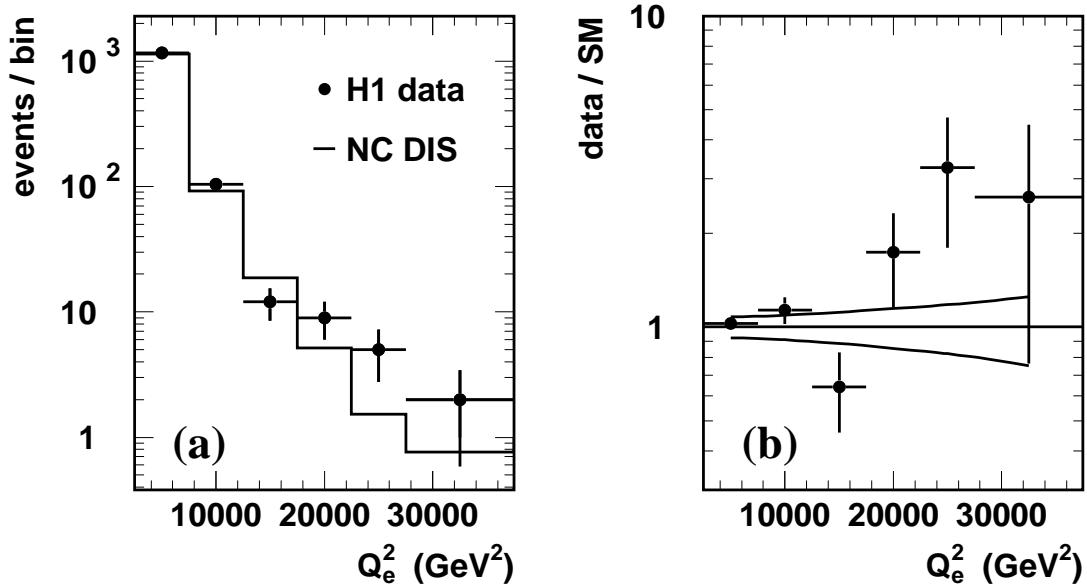} 
    \caption[]{ \label{fig:q2plotnc}
      { \small  (a) $Q^2_e$ distribution of the selected NC DIS candidate
                events for the data ($\bullet$ symbols) and for standard
                NC DIS expectation (histogram);
                (b) ratio of the observed and expected (from
                NC DIS) number of
                events as a function of $Q^2_e$;
                the lines above and below unity specify the
                $\pm 1\sigma$ levels determined using the combination
                of statistical and systematic errors of
                the DIS expectation. }}
 \end{center}
\end{figure}

We then consider the $Q^2$ information. 
Figure~\ref{fig:q2plotnc}a shows the measured $Q^2_e$ distribution in 
comparison with the expectation from standard NC DIS processes.
Also shown in Fig.~\ref{fig:q2plotnc}b is the ratio of the observed $Q^2_e$
distribution to the NC DIS expectation.
%
%
%
The errors resulting from the convolution of the
systematic errors and the statistical error of the Monte Carlo sample
are correlated for
different $Q^2_e$ bins and are indicated in Fig.~\ref{fig:q2plotnc}b
as lines above and below unity joining the $\pm 1\sigma$ errors
evaluated at the centre of each bin.
These errors are dominated by the uncertainty in the electromagnetic
energy scale of the calorimeter and vary between $7.7\%$ at low
$Q^2_e$ and $25\%$ at the highest values of $Q^2_e$.
The NC DIS expectation agrees well with the data for
$Q^2_e \lsim 10000 \GeV^2$ while at larger $Q^2_e$, deviations are
observed, with a slight deficit around $15000 \GeV^2$ and a number of 
observed events at $Q^2 \, \gsim \, 15000 \GeV^2$ in excess of the NC DIS 
expectation.
For $Q^2 > 15000 \GeV^2$, 22 events are observed while
$14.1 \pm 2.0$ are expected from standard NC DIS.

\subsection{Charged Current Deep-Inelastic-like Signatures}
\label{sec:selectcc}

 \subsubsection{Event selection and kinematics}
\label{sec:selectcckine}

The inclusive {\bf selection of CC DIS-like events} requires:
\begin{enumerate}
  \item no $e^{\pm}$ candidate with $E_T > 5 \GeV$ found in
        the LAr calorimeter;
  \item the total missing transverse momentum 
         $P_{T,miss}  > 30 \GeV$.
\end{enumerate}
These cuts eliminate the photoproduction and NC DIS background.
To deal with specific background sources to CC DIS, it is required that 
there be no isolated track with $P_T > 10 \GeV$ found within the 
angular range $10^{\circ} \le \theta \le  145^{\circ}$.
This reduces the remaining contamination to $< 0.3 \%$ from misidentified 
NC DIS events where the positron has been scattered through a crack of the 
calorimeter and also suppresses eventual background from single $W$ boson, 
while causing negligible efficiency losses for the CC DIS selection.

The $Q^2$, $y$ and $M$ are calculated using the Jacquet-Blondel 
ansatz~\cite{JACQUET} by summing over all measured final state 
hadronic energy deposits using:
$$ M_h = \sqrt{\frac{Q^2_h}{y_h}}, \;\;\;\;
   Q^2_h= \frac{P^2_{T,miss}}{1-y_h},\;\;\;\;
   y_h=\frac{\sum \left(E-P_z\right)}{2E_e^0}.\;\;\;\; $$
This method will henceforth be called the hadron method ($h$-method).

In addition to the cuts (1) and (2) above, the analysis is restricted 
to the kinematic domain $Q^2_h >  2500 \GeV^2$ and $y_h < 0.9$.
The resolutions in both $M_h$ and $Q^2_h$ degrade with increasing $y$
since both $\delta M_h / M_h$ and $\delta Q^2_h / Q^2_h$ behave as
$1/(1-y_h)$ for $y_h \sim 1$. 
Hence the high $y_h$ domain is excluded.
Throughout the remaining domain, the CC trigger efficiency is $96.5 \pm 2 \%$.
For $LQ \rightarrow \nu + q$ events, these selection criteria 
ensure typical efficiencies varying between $\simeq 32 \%$ and 
$\simeq 79 \%$ for LQ masses ranging in $75$ to $250 \GeV$.

Following this CC selection, $213$ DIS event candidates are accepted in
good agreement with the standard CC DIS expectation of $199.1 \pm 11.5$
events. 

\subsubsection{Comparison with Standard Model expectation}
\label{sec:DISCCsample}

\begin{figure}[htb]
   \begin{center}
     \epsfxsize=1.0\textwidth
     \epsffile{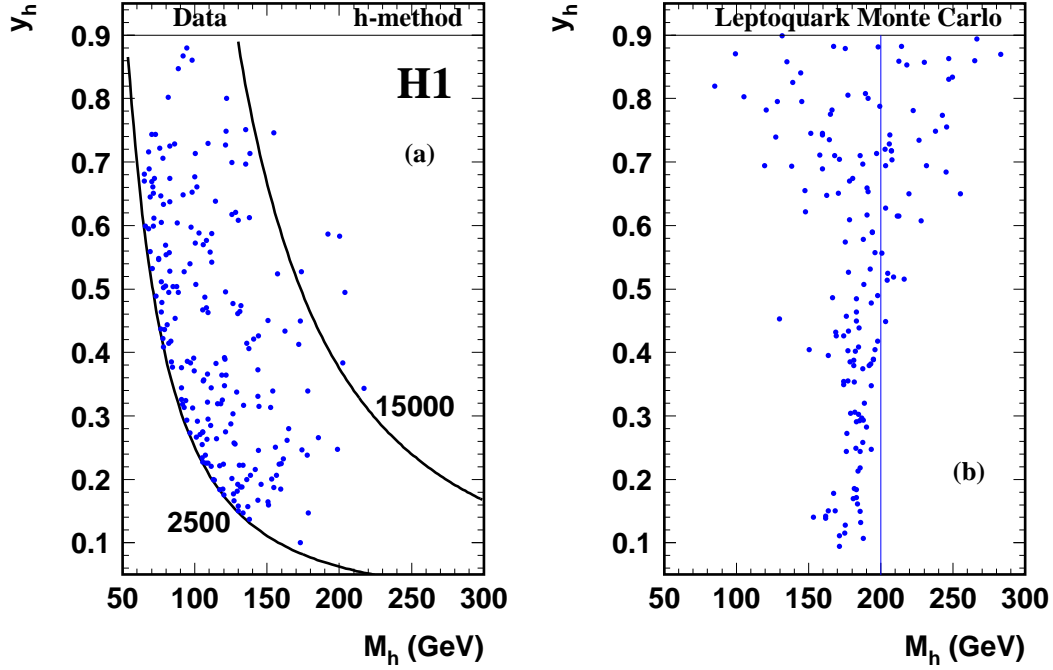}
      \caption
         {\small \label{fig:scattercc}
                 Kinematics in the $y_h$ - $M_h$ plane of
                (a) the selected CC DIS candidate events from H1 data
                    (two isocurves at $Q^2 = 2500$ and $15000 \GeV^2$
                    are plotted as full lines);
                (b) a scalar leptoquark resonance
                    of mass $M_{LQ} = 200 \GeV$ decaying into $\nu + q$,
                    for a coupling $\lambda = 0.05$. }
 \end{center}
\end{figure}

Figure~\ref{fig:scattercc}a shows the two dimensional distribution of
$y_h$ against $M_h$ for the CC candidates. 
Signal Monte Carlo events coming from the decay of a $200 \GeV$ narrow
scalar resonance into $\nu + q$ are shown in the same plane
in Fig.~\ref{fig:scattercc}b, where the degradation of the resolution
in $M_h$ at high $y_h$ is clearly visible.
This resolution on the LQ mass is of about $10 \%$.
While the relative calibration of the hadronic scale is known at
the $2\%$ level (as discussed in section~\ref{sec:detector} and controlled
comparing real and simulated NC DIS events), the energy scale procedure which
relies on transverse momentum balance does not attempt to correct on
average in DIS events the measured $M_h$ to a ``true" value. 
The measured $M_h$ underestimates the resonance mass systematically by
$\simeq 6 \%$ and this shift will be taken into account in deriving LQ 
results.
%

\begin{figure}[htb]
  \begin{center}
    \epsfxsize=1.0\textwidth
    \epsffile{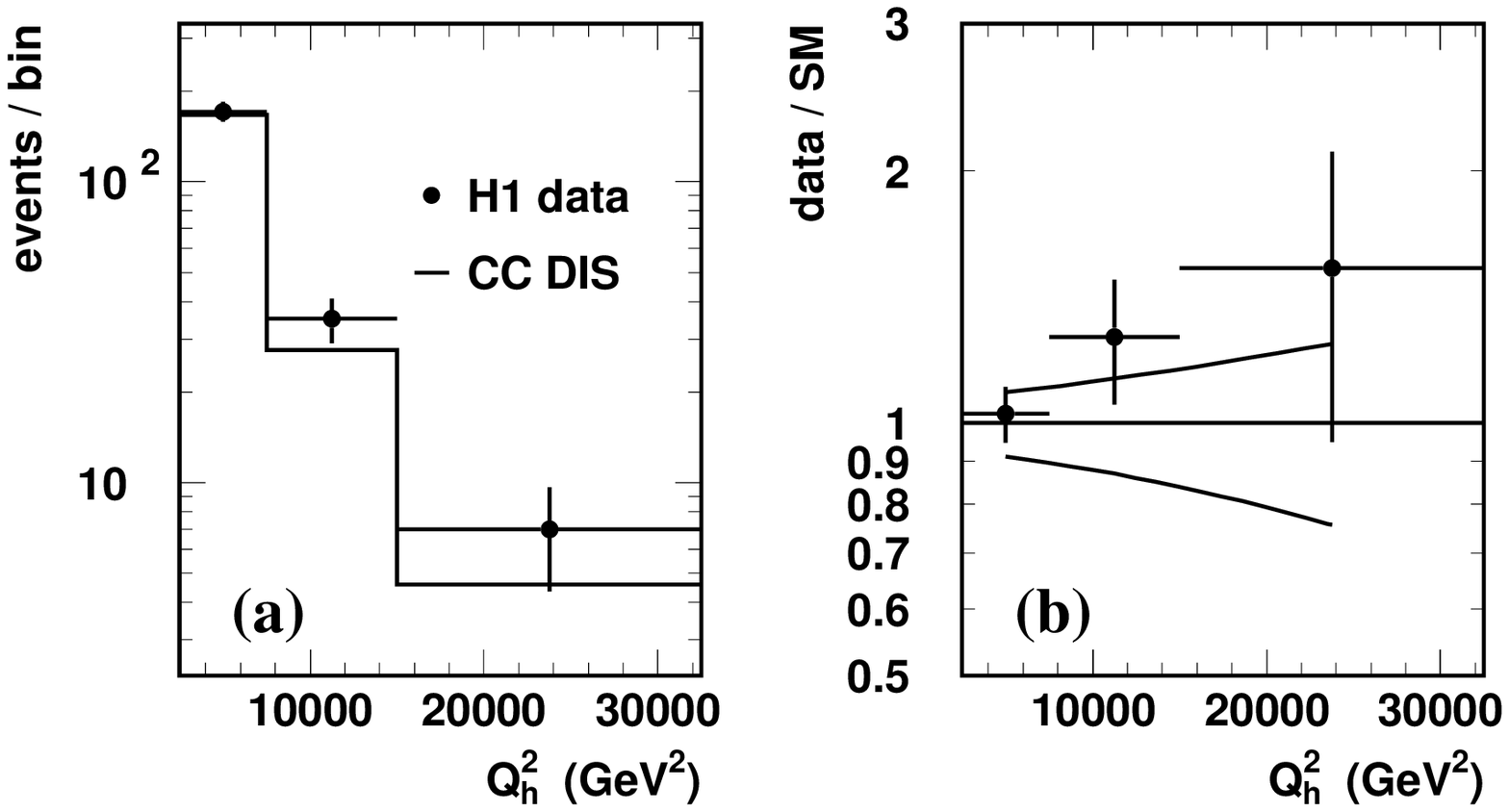}
    \caption[]{ \label{fig:q2plotcc}
     { \small   (a) $Q^2_h$ distribution of the selected CC DIS candidate
                events for the data ($\bullet$ symbols) and for
                standard CC DIS expectation (histogram);
                (b) ratio of the observed and expected (from CC
                DIS) number of
                events as a function of $Q^2_h$;
                the lines above and below unity specify the
                $\pm 1\sigma$ levels determined using the combination
                of statistical and systematic errors of
                the DIS expectation. }}
 \end{center}
\end{figure}

Figures~\ref{fig:q2plotcc}a and~\ref{fig:q2plotcc}b show for the CC selection 
the measured $Q^2_h$ distribution in comparison with the standard CC DIS
expectation.
As can be seen in Fig.~\ref{fig:q2plotcc}b, the systematic
errors are relatively large and dominated by the uncertainty
on the hadronic energy scale of the calorimeter.
In the kinematic region $Q^2_h>15000 \GeV^2$, there are $N_{obs}=7$
observed events compared with an expectation of $4.84 \pm 1.42$
from standard CC DIS.
%
%

\subsection{High {\boldmath $P_T$} Muon Signatures}
\label{sec:selectmu}

For {\bf LQs possessing a coupling to a second generation lepton}, 
leading to $\mu+q$ final states, we start from a CC-like selection in the 
kinematic domain $Q^2_h > 1000 \GeV^2$ and $P_{T,miss} > 25~\GeV$, and then 
require:
\begin{enumerate}
  \item an identified isolated muon with transverse momentum
        $P_{T,\mu} > 10 \GeV$ and within polar angular range 
        $10^{\circ} < \theta_{\mu} < 145^{\circ}$,
        where the momentum and angle are determined
        by the associated track. There should be no other charged
        track linked to the primary interaction vertex within the
        pseudorapidity-azimuthal isolation cone centred on the
        $\mu$ track of opening 
        $\sqrt{ (\Delta \eta_{\mu})^2 + (\Delta \phi_{\mu})^2 } = 0.5$; 
  \item at least one jet found in the angular range 
        $7^{\circ} < \theta < 145^{\circ}$,
        with a transverse momentum $P_T > 15 \GeV$,
        using the cone algorithm mentioned in~\ref{sec:selectnckine}.
\end{enumerate}
The muon identification combines inner tracking and calorimetric
information.
Within the $\mu$ isolation cone, at least $1 \GeV$ must be visible
in the calorimeters.
Restricting to the LAr calorimeter, this energy should be
smaller than one third of the $\mu$ track momentum.
Less than $5 \GeV$ should be seen in the conical
envelope between
$0.5 < \sqrt{ (\Delta \eta_{\mu})^2 + (\Delta \phi_{\mu})^2 } < 1.0$.
The centroid of the energy deposits in the calorimeters within
the $\mu$ isolation cone should not be in the LAr electromagnetic section.
With these identification criteria, muons are found with a typical 
efficiency of $\simeq 85 \%$ over most of the angular range.

Only four $\mu + jet$ events are observed after applying
these basic requirements. These four ``outstanding'' events are 
amongst\footnote{
  The event labelled $\mu_3$ in~\cite{H1MUEV}, in which both
  an isolated muon and the scattered positron are identified,
  does not fulfil the first requirement 
  (see section~\ref{sec:selectcckine})
  we apply to select CC-like events. }
the high $P_T$ lepton events discussed in~\cite{H1MUEV}.
With these selection criteria $0.60 \pm 0.10$ events are expected from
SM processes, coming mainly from $W$ production
and inelastic photon-photon interactions 
$\gamma \gamma \rightarrow \mu^+ \mu^-$.

For the $2 \rightarrow 2$ body LQ induced processes, where the
final state consists of only the muon and the scattered quark,
the energy-momentum conservation relates the polar angle
$\theta_l$ of the muon to the measurement of the hadronic final
state, by~:
$$ \tan \frac{\theta_l}{2} = \frac { 2 E^0_e - \sum (E-p_z) }
                    {\sqrt{ (\sum E_{x})^2 + (\sum E_{y})^2 }}
    \qquad . $$
The selection of $\mu + jet$ candidates induced by LQ decay or
$u$-channel exchange requires that~:
\begin{enumerate}
 \renewcommand{\labelenumi}{(i)}
  \item the polar angle $\theta_{\mu} $ of the high $P_T$ 
        isolated track agrees with $\theta_l$ within $30^{\circ}$;
  \renewcommand{\labelenumi}{(ii)}
  \item the track and the hadronic flow are back-to-back
        $(\Delta \phi_{\mu - h} > 170 ^{\circ})$
        in the plane transverse to the beam axis.
\end{enumerate}

\begin{figure}[tb]
  \begin{center}
  \vspace{-1.0cm}
   \epsfxsize=0.6\textwidth
   \epsffile{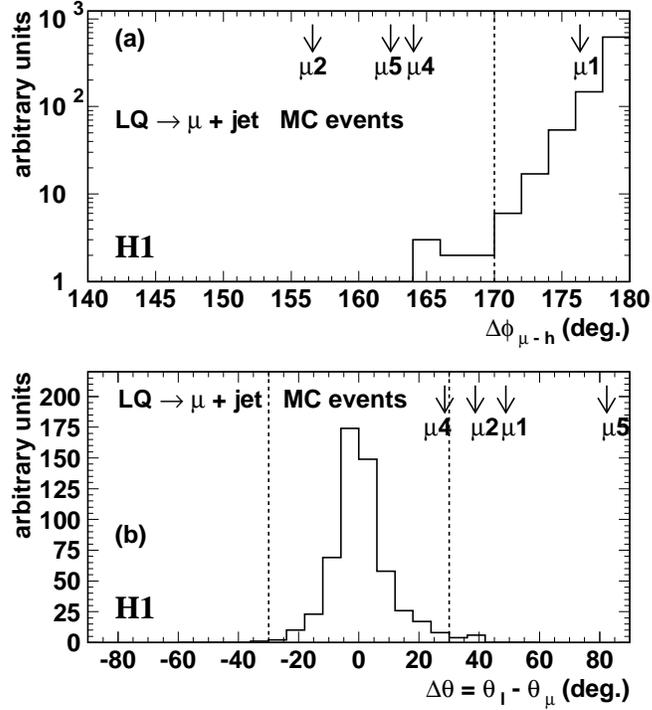}
   \caption[]{ \label{fig:lqmu}
      {\small (a) Azimuthal opening between the track associated to
                  the high $P_T$ muon in $e + p \rightarrow \mu + jet$
                  events and the hadronic energy flow;
              (b) polar angle opening between the muon track and the
                  final state lepton angle predicted from the hadronic
                  energy flow for a $2 \rightarrow 2$ body process.
              Arrows indicate the $\Delta \phi$ and $\Delta \theta$
              values for the ``outstanding" events reported
              in~\cite{H1MUEV} 
              and labelled as therein.
              Histograms represent the expected
              distributions of these variables for
              $eq \rightarrow \mu q'$ LQ Monte Carlo events.}}
  \end{center}
 \end{figure}

As seen in Fig.~\ref{fig:lqmu}, these criteria induce a minute
efficiency loss for a LQ signal simulation. The actual loss
was estimated to be $\simeq 5 \%$ using real NC-like data by searching
for the charged track angles associated with the positron candidate
as deduced using the hadronic energy flow.
The above requirements lead to typical efficiencies to select
$LQ \rightarrow \mu + q$ events which vary between $\simeq 30 \%$
and $\simeq 60 \%$ for LQ masses ranging in $75$-$250 \GeV$.
For LQ masses far above the kinematic limit ($M_{LQ} \gg \sqrt{s_{ep}}$),
the distribution of the polar angle of the final state lepton
depends on the LQ spin and fermion number (this latter affecting
the relative contributions of the $s$- and $u$-channels).
With the above selection criteria the efficiency to select
$\mu + jet$ events induced by very heavy LQs varies between
$\simeq 20 \%$ and $\simeq 33 \%$ depending on the LQ quantum
numbers, but independently on the LQ mass.

We observe no candidate satisfying the LQ $\mu+jet$ selection
while $0.12 \pm 0.05$ are expected from SM processes
(mainly from inelastic photon-photon interactions).
In particular, the $e^+ p \rightarrow \mu^+ X$ outstanding events
discussed in~\cite{H1MUEV} fail significantly the kinematic
constraints for LQ induced $ e q \rightarrow \mu q' $ processes
as can be seen in Fig.~\ref{fig:lqmu}.
%

\subsection{High {\boldmath $P_T$} Tau Signatures}
\label{sec:selectau}

For {\bf LQs possessing a coupling involving a third generation lepton} 
leading to
$\tau + q$ final states, the analysis is restricted to hadronic
decays of the $\tau$~\footnote{
The $\tau^+ \rightarrow \mu^+ \nu_{\mu} \bar{\nu}_{\tau}$ 
channel is covered 
implicitly by the above $\mu + q$ search since the $\tau$ decay products 
are strongly boosted in the $\tau$ direction; the
$\tau^+ \rightarrow e^+ \nu_e \bar{\nu}_{\tau}$ channel is not 
covered here.}.
The identification of ``pencil-like'' jets induced by hadronic decays
of $\tau$ requires that the jet invariant mass satisfies
$M_{jet} \le 7 \GeV$ and has a low multiplicity, namely
$1 \leq N_{tracks} \leq 3$, $N_{tracks}$ being the number of vertex
fitted tracks with $P_T > 0.15 \GeV$ in the jet identification cone.
The jet mass $M_{jet}$ is here calculated as the invariant mass of
all energy deposits associated to the jet, each of those being treated
as a massless object.

Inclusive $\tau$ + jet signatures, where the $\tau$ lepton
decays into hadrons, are selected by requiring that~:
\begin{enumerate}
  \item no $e^{\pm}$ candidate with $E_T > 5 \GeV$ is found in
        the LAr calorimeter;
  \item two jets are found in the angular range
        $7^{\circ} < \theta < 145^{\circ}$
        using the cone algorithm mentioned in~\ref{sec:selectnckine}
        with a transverse momentum $P_T > 30 \GeV$;
        one of these jets must satisfy the ``loose'' $\tau$-jet
        identification criteria described above;
  \item there is at most a small amount of energy deposited in the
        backward calorimeter, $E_{back} < 7.5 \GeV$;
  \item the impact point of the ``leading track'' associated
        to the $\tau$-jet candidate at the inner surface
        of the LAr calorimeter must be at least $2^{\circ}$ apart in azimuth
        from each of the eight $\phi$ cracks of the LAr.
        The ``leading track'' is, among the vertex fitted tracks found
        in the jet identification cone, the one which has the highest
        momentum projected on the jet axis;
  \item the leading track associated to the $\tau$-jet
        candidate carries a
        large enough fraction $E_{track}/E_{jet} > 10\%$ of the jet
        energy $E_{jet}$;
  \item the fraction of the $\tau$-jet candidate energy carried by the
        leading track and the energy deposition fraction $f_{em}$ in
        the LAr electromagnetic section of the $\tau$-jet are such that
        $$f_{em} + E_{track}/E_{jet} < 1.5 \qquad ; $$ 
  \item
        the energy deposits of the jet should present 
        an important longitudinal dispersion
        $R_L = \sqrt{ \langle l^2 \rangle - \langle l \rangle^2} > 7 \cm $
        where, for each deposit, $l$ is the distance from the impact point
        at the calorimeter surface along the jet axis; 
  \item at most two vertex fitted tracks are found in the
        azimuthal hemisphere containing the $\tau$-jet candidate.
\end{enumerate}
%
 \begin{figure}[hbt]
  \begin{center}
 \epsfxsize=0.7\textwidth
  \epsffile{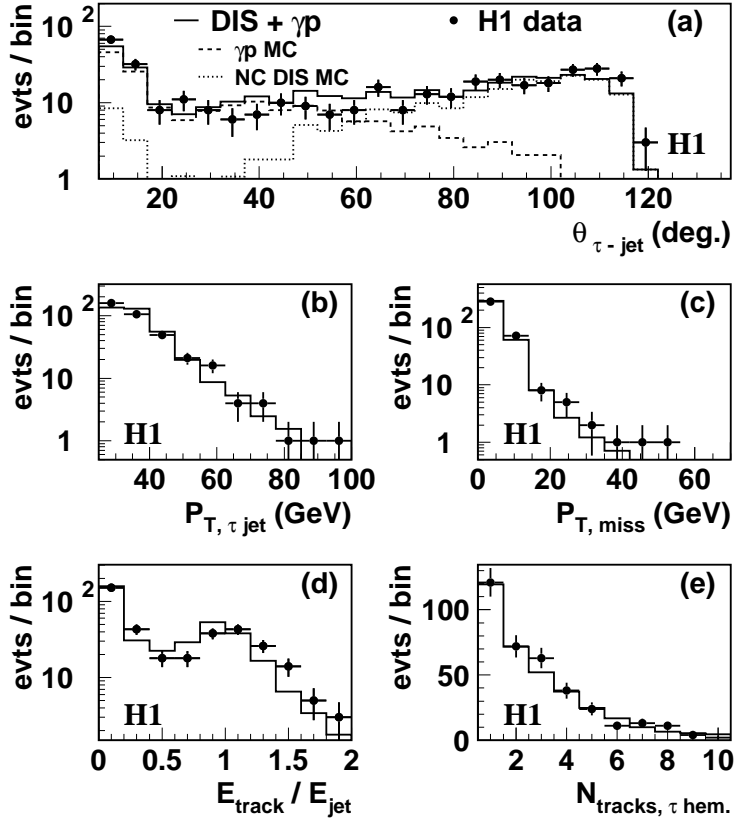}
   \caption[]{ \label{fig:tauctrl}
      {\small Distributions of (a) the polar angle
              $\theta_{\tau-jet}$ of the $\tau$-jet candidate;
              (b) the transverse momentum of the $\tau$-jet
                  candidate;
              (c) the total missing transverse momentum;
              (d) the fraction of the energy of the $\tau$-jet
              candidate
              carried by
              the leading track and (e) the number of tracks in
              the azimuthal hemisphere containing the $\tau$
              candidate. Symbols correspond to data events and
              histograms to SM simulation. }}
   \end{center}
  \end{figure}
%
 \begin{figure}[htb]
 \begin{center}
 \vspace*{-1cm}
 \begin{tabular}{p{0.58\textwidth}p{0.37\textwidth}}
     \hspace*{-0.5cm}\raisebox{-200pt}{
     \mbox{\epsfxsize=0.65\textwidth
         \epsffile{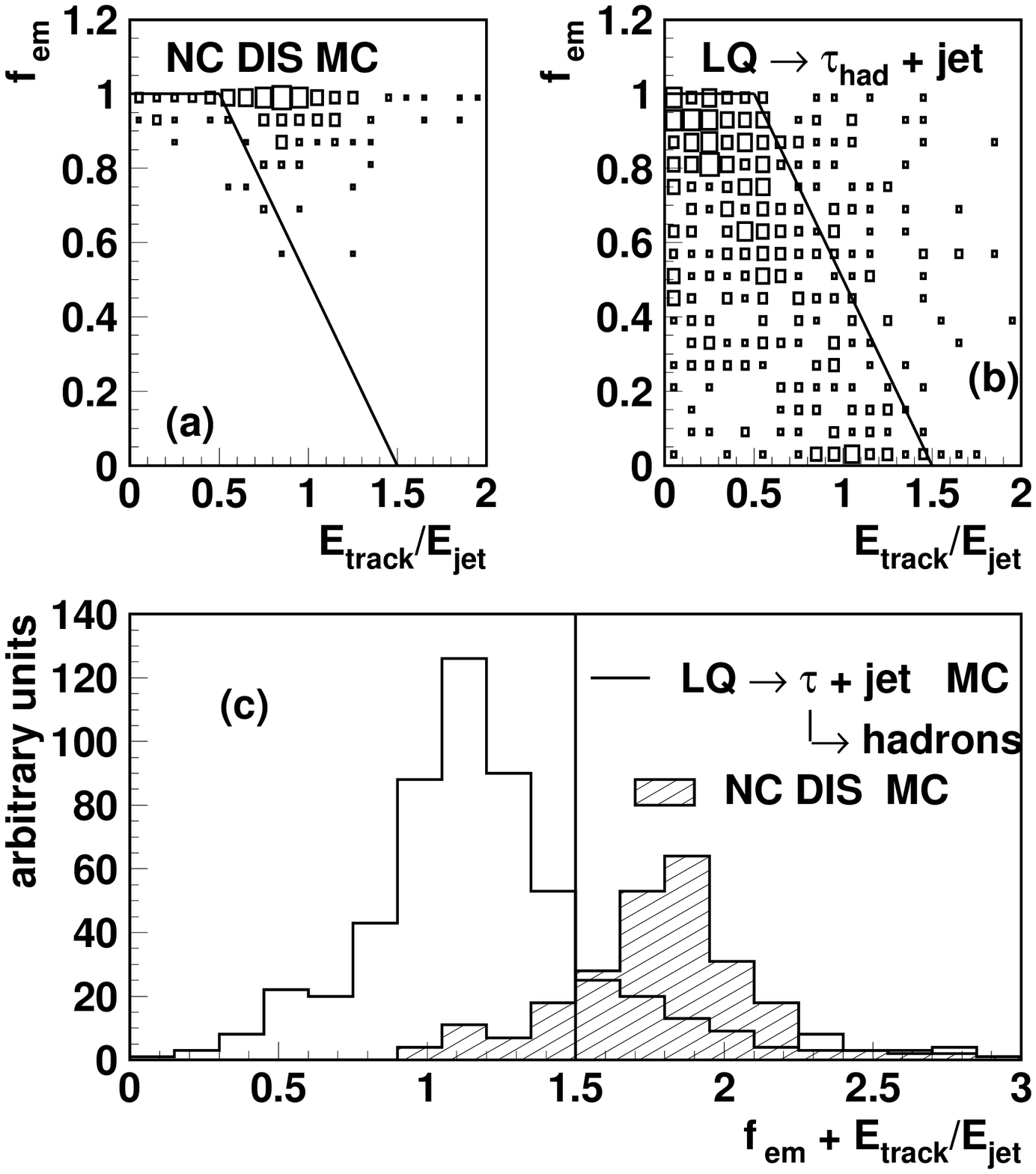}}}
 &
         \caption
         {\small \label{fig:disreduc}
              Correlation between the electromagnetic fraction
              of the $\tau$-jet candidate and the fraction of
              the jet energy carried by the leading track
              for (a) NC DIS and (b)
              $LQ \rightarrow \tau + q$ simulated events;
              (c) shows the sum of these variables and the cut
              applied. 
              Events where the track associated to the $\tau$-jet
              candidate is at less than $2^{\circ}$ in
              azimuth from a $\phi$ crack of the calorimeter
              have been excluded. }
 \end{tabular}
 \end{center}
\end{figure}

Cuts (1) to (3) ensure a preselection of $\tau$ + jet candidate events at high
$P_T$; low $Q^2$ NC DIS is suppressed by cut (3). At this stage,
375 events are selected in the data while 365.7 are expected from
NC DIS and $\gamma p$ processes.
Figure~\ref{fig:tauctrl}a shows, for these selected candidates, the
distribution of the polar angle $\theta_{\tau-jet}$
of the $\tau$-jet candidate.
Photoproduction background, where a low multiplicity jet can fake
a $\tau$-jet, is seen to contribute mainly at low
$\theta_{\tau-jet}$ in contrast to NC DIS background
arising when the scattered positron has not been identified.
NC DIS contamination at low values of $\theta_{\tau-jet}$ is due to events 
where the positron has been scattered at large angle through a crack of 
the calorimeter, and where the ``current jet'' has been identified as a 
high $P_T$ $\tau$-jet.
Figures~\ref{fig:tauctrl}b and c show the distributions of
the transverse momentum of the $\tau$-jet and of the whole final state.
Figure~\ref{fig:tauctrl}d shows the ratio of the energy of the
leading track associated with the $\tau$-jet candidate, to the
$\tau$-jet energy.
The observed distribution of this fraction $E_{track} / E_{jet}$ 
is shifted by $ 9 \%$ compared to the one expected from
the simulation. 
This shift is taken into account as a systematic
error on the quantity $E_{track} / E_{jet}$, propagated
when estimating the uncertainty on the SM expectation.
On Fig.~\ref{fig:tauctrl}e the number of tracks found in the azimuthal 
hemisphere containing the $\tau$-jet candidate is shown to be well 
described by the simulation.

NC DIS and $\gamma p$ backgrounds are further reduced
by requirements (4) to (8).
Cut (4) avoids regions close to $\phi$ cracks of the calorimeter
and hence suppresses NC DIS background.
Cut (5) efficiently reduces $\gamma p$ background where a low multiplicity
jet could fake a $\tau$-jet.
Cut (6) ensures a powerful suppression of the remaining NC DIS background 
events where the scattered positron has not been identified.
It exploits the fact that the electromagnetic fraction of a $\tau$-jet
is high mainly when the $\tau$ lepton decays to $n \pi^0 + X$,
which generally implies a small value for the ratio $E_{track}/E_{jet}$.
On the contrary, for NC DIS events where the positron has been
misidentified because its shower has a substantial leakage into the 
hadronic section of the calorimeter, this ratio 
$E_{track}/E_{jet}$ is expected to peak at one.
Figure~\ref{fig:disreduc} illustrates this property of $\tau$ jets
and shows how cut (6) discriminates the searched signal
from the DIS background.
Remaining NC DIS events are further removed by cut (7).
The more stringent requirement on the track multiplicity
imposed by cut (8) further reduces the remaining $\gamma p$ contamination.
This cut takes into account the fact that two reconstructed
tracks can be associated to a single charged particle, 
especially when scattered in the forward region.

The efficiency to identify $\tau$ leptons decaying
hadronically using the above criteria is $\simeq 25\%$.

We observe 21 events satisfying the above inclusive $\tau$ + jet 
requirements, which agrees well with the mean expectation of 
$25.4 \pm 4.3$ events coming from NC DIS and $\gamma p$ processes. \\

\begin{figure}[h]
 \begin{center}
 \begin{tabular}{p{0.58\textwidth}p{0.37\textwidth}}
     \hspace*{-0.5cm}\raisebox{-200pt}{
     \mbox{\epsfxsize=0.65\textwidth
     \epsffile{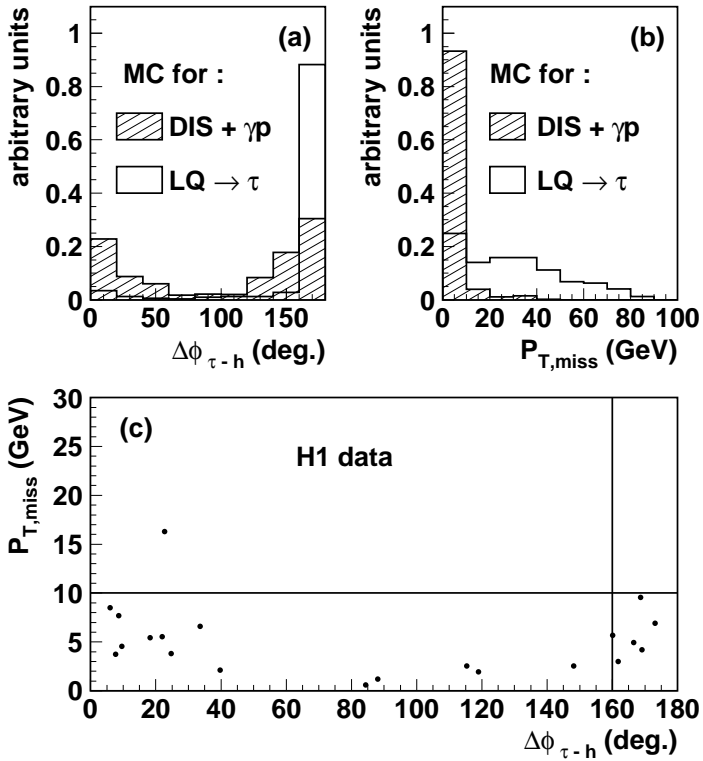}}}
 &
         \caption
         {\small \label{fig:lqtau}
          Distributions of (a) the azimuthal angular opening
          $\Delta \phi_{\tau-h}$ between the $\tau$-jet
          candidate and the hadronic energy flow;
          (b) the missing transverse momentum $P_{T,miss}$.
          White and hatched histograms correspond to the
          SM expectation and the LQ induced $\tau$+jet signal
          respectively.
          (c) Correlation of these two variables for the
          data events satisfying cuts (1) to (8). 
          The full lines show the cuts applied, which retain
          the upper-right part of this plane. }
 \end{tabular}
 \end{center}
\end{figure}

For the $\tau + X$ channel, additional cuts relevant for the specific 
LQ search are applied~:
\begin{enumerate}
 \renewcommand{\labelenumi}{(i)}
 \item a total transverse momentum $P_{T,miss} > 10 \GeV$ ;
 \renewcommand{\labelenumi}{(ii)}
 \item the $\tau$-jet candidate is at
       $\Delta \phi_{\tau-h} > 160^{\circ}$
       from the total hadronic flow.
\end{enumerate}
Both these cuts exploit the fact that neutrino(s) emerging from the
$\tau$ decay
are collimated with the $\tau$-jet direction, and that the other jet
in LQ induced $\tau + {\mbox{jet}}$ events is formed by the
fragmentation of a quark and thus should not
contribute to the missing transverse momentum.
Moreover, cut (i) efficiently reduces the remaining
NC DIS contamination.
The correlation between the $P_{T,miss}$ and the $\Delta \phi_{\tau-h}$
for the 21 data events which satisfy the criteria (1) to (8)
listed above
is shown in Fig.~\ref{fig:lqtau}, together
with the expected distributions of $\Delta \phi_{\tau-h}$
and $P_{T,miss}$ for LQ induced $\tau + {\mbox{jet}}$
Monte Carlo events.

The final selection efficiency on the LQ $\tau + q$ signal is
$\simeq 10 \%$ for LQ masses of $100 \GeV$ and reaches a plateau
at $\simeq 25 \%$ above $200 \GeV$.
The small efficiency at low masses is mainly due to the requirement
of two high $P_T$ jets.
For LQ masses far above the kinematic limit, this efficiency varies
between $\simeq 8 \%$ and $\simeq 12 \%$ depending on the LQ spin
and fermion number.

We observe no candidate satisfying the additional cuts designed specifically 
for LQ induced processes involving third generation leptons for an 
expectation of $0.77 \pm 0.30$ event from misidentified electrons in 
NC DIS processes.

It should be noted that the absence of $\tau$ + jet candidates accompanied 
by a large $P_{T,miss}$, despite the relatively loose $\tau$-jet 
requirements, makes it rather unlikely for the $\mu$'s in the 
$e^+ p \rightarrow \mu^+ X$ events~\cite{H1MUEV} to originate
from a jet fluctuating in one single particle.
Conversely, these muon events fail significantly the kinematic
requirements of a two body decay $LQ \rightarrow q + \tau$ followed
by a $\tau \rightarrow \mu^+ \nu_{\mu} {\bar \nu_{\tau}}$ decay
since the final state $\mu^+$ would appear boosted in the $\tau$
direction and the $\Delta \theta$-$\Delta \phi$ restrictions of
Fig.~\ref{fig:lqmu} would still apply.

\section{Constraints on First Generation Leptoquarks}
\label{sec:lqfirst} 

\subsection{The Leptoquark Specific Angular Cut }
\label{sec:ycut}

We first consider LQs possessing couplings to $e-q$ or $e-\bar{q}$ 
pairs only. 

\begin{figure}[htb]
  \begin{center}
     \mbox{\epsfxsize=0.8\textwidth
           \epsffile{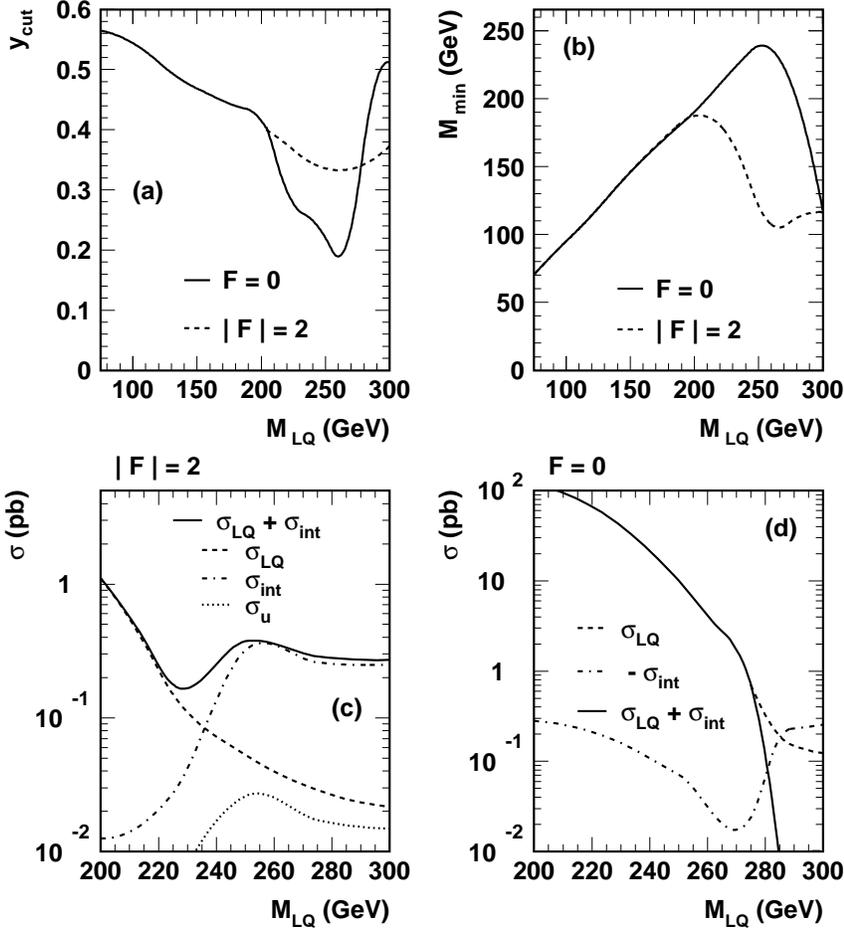}}
  \end{center}
 \caption[]{ \label{fig:cuts_inter}
 {\small Lower cut applied (a) on $y_e$ and (b) on $M_e$ to enhance the 
         signal significance for a $F=0$ (full line) and a $\mid F \mid = 2$ 
         (dashed line) scalar leptoquark, for NC DIS-like final states;
         contribution $\sigma_{LQ}$ of the LQ induced 
         $e^+ p \rightarrow e + q + X$ processes
         ($s$- and $u$-channels summed) 
         and of their interference $\sigma_{int}$ with SM DIS 
         for
         (c) $S_{0,R}$ ($\mid F \mid = 2$) and (d) $S_{1/2,L}$ ($F=0$) 
         leptoquark, with $\lambda=0.5$. 
         In (c), the contribution of the $u$-channel LQ exchange alone
         is also represented (dotted line).
         Each contribution has been integrated over
         the phase space allowed by the mass dependent $M_e$ and $y_e$ 
         cuts applied in the analysis. }}
\end{figure}

In order to enhance the significance of a possible LQ signal over
the NC DIS background remaining after the selection requirements
described in section~\ref{sec:selectnckine}, the specific angular distributions
for LQ induced processes are exploited.
For scalar or vector LQs with either $F=0$ or $\mid F \mid = 2$, distinct mass 
dependent lower $ y_e > y_{cut}$ cuts have been optimized using Monte Carlo 
generator programs to maximize the signal significance.
This has been achieved by finding the best compromise between the 
efficiency loss on the LQ signal and the important background reduction.
Thus, with increasing LQ mass, the $y_{cut}$ decreases together with the
NC DIS expectation.
However, for LQ masses close to the kinematic limit, the mass spectrum is 
highly distorted towards low values (as explained in 
section~\ref{sec:pheno}). Thus, a higher $y_{cut}$ will be needed to
enhance the signal significance of such a high mass LQ observed at
$M_e \ll M_{LQ}$ where a larger NC DIS background is expected.
The evolution of $y_{cut}$ as a function of the LQ mass is represented 
in Fig.~\ref{fig:cuts_inter}a, for the example of $F=0$ and $\mid F \mid = 2$ 
scalar LQs.
%
%
%
For a scalar $F=0$ LQ, the $y_{cut}$ monotonously decreases from 
$\simeq 0.6$ around $60 \GeV$ to $\simeq 0.2$ around $260 \GeV$, 
and then rises up to $\simeq 0.5$ at $300 \GeV$.
The behaviour is similar for a scalar $\mid F \mid = 2 $ LQ, but the effect 
of the distortion of the mass spectrum is significant
already at $ \simeq 200 \GeV$. 
A similar description holds for vector LQs but smaller values of
the $y_{cut}$ are obtained, because of the $(1-y)^2$ shape of their $y$ 
spectra.

\subsection{The Mass Spectra}
\label{sec:lqmass}

%
\begin{figure}[htb]
  \begin{center}
  \begin{tabular}{cc}
     \hspace*{-0.9cm}\mbox{\epsfxsize=0.55\textwidth
        \epsffile{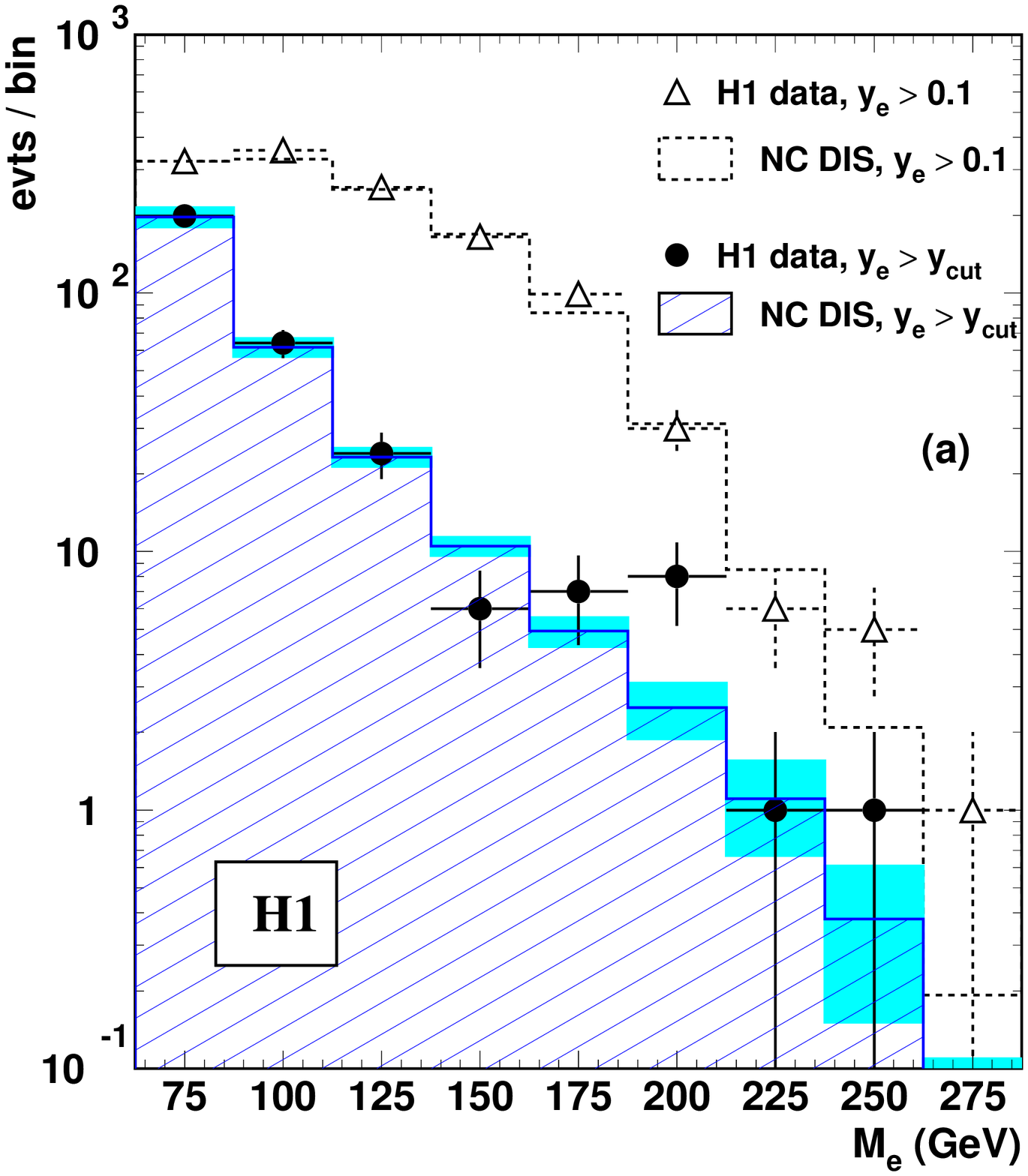}}
   &
     \hspace*{-0.8cm}\mbox{\epsfxsize=0.55\textwidth
     \epsffile{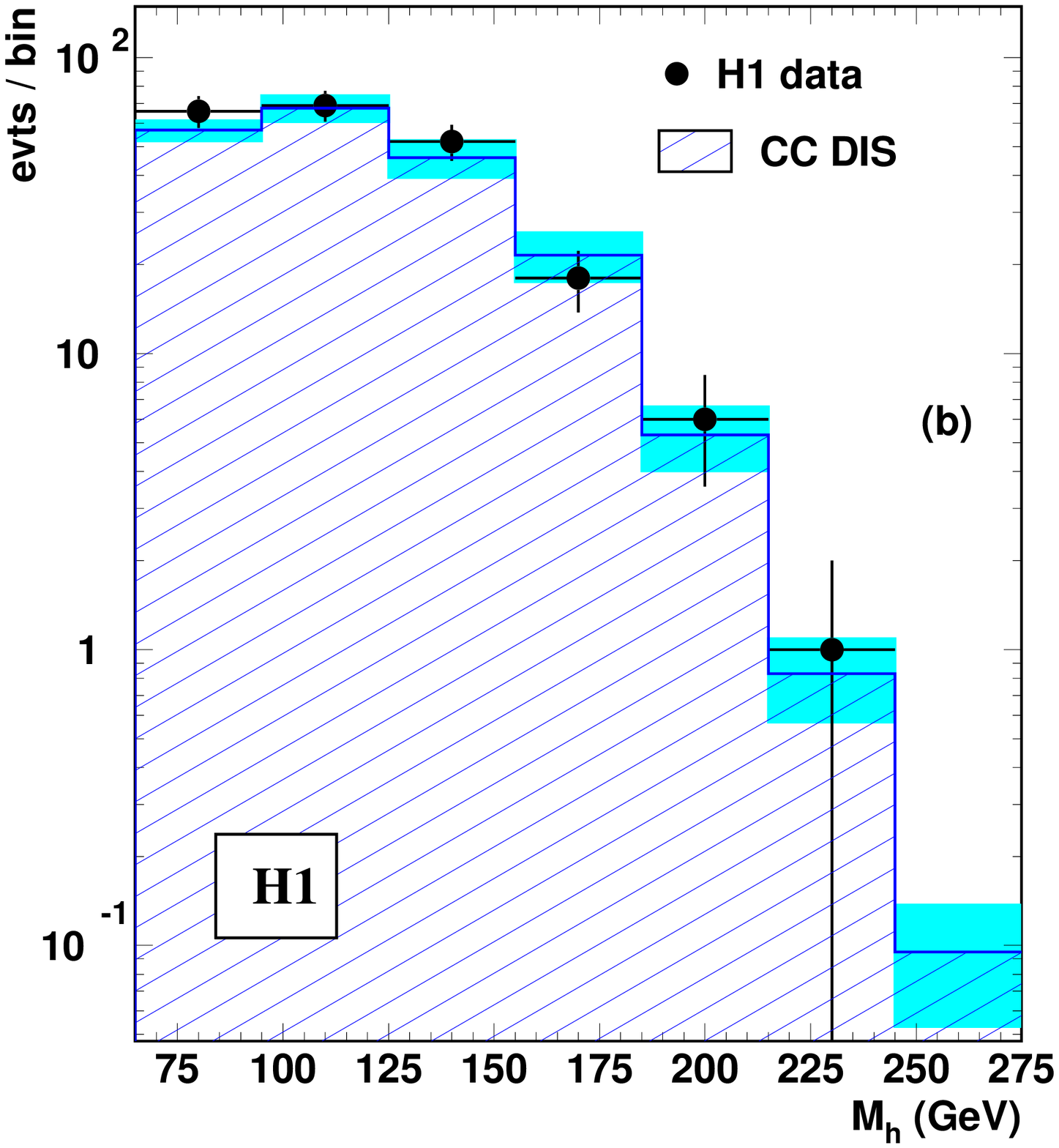}}
  \end{tabular}
  \end{center}
 \caption[]{ \label{fig:dndmlq}
 {\small Mass spectra for (a) NC DIS-like and (b) CC DIS-like final states
         for data (symbols) and DIS expectation (histograms).
         In (a), the NC DIS-like comparison is shown before (open triangles,
         white histogram) and after (closed dots, hatched histogram)
         a $y$ cut designed to maximize the significance of an eventual
         scalar LQ signal.
         The greyed boxes indicate the $\pm 1 \sigma$ band combining the 
         statistical and systematic errors of the NC and CC DIS 
         expectations. }}
\end{figure}
The comparison of the measured mass spectrum with SM predictions
is shown in Fig.~\ref{fig:dndmlq}a for the NC DIS analysis.
The measured and expected NC spectra are seen before and after applying the
mass dependent $y_e$ cut relevant for a $F=0$ scalar LQ. 
After this $y_e$ cut, we observe 310 events in the mass range 
$M_e > 62.5 \GeV$ while $301.2 \pm 22.7$ are expected.
The observed mass spectrum is seen to be well described by the
SM expectation, with nevertheless a slight excess in the
mass range $ 200 \GeV \pm \Delta M / 2$ with $\Delta M = 25 \GeV$,
due to the same 8 events already discussed in 
section~\ref{sec:DISNCsample} (the cut optimal for $F=0$ LQs is
$y_{cut} \simeq 0.4$ for $M \simeq 200 \GeV$).
For CC DIS-like final states, the mass distribution of the 213 events
fulfilling the requirements described in section~\ref{sec:selectcckine} is seen
in Fig.~\ref{fig:dndmlq}b to be in good agreement with the SM prediction.

\subsection{The Limits Derivation}
\label{sec:limimeth}

Assuming that the observed excess of events in the NC DIS-like channel
is due to a statistical fluctuation (or either, formally,
that the event sample
contains at most two components, an unknown LQ signal and a
known expectation from NC DIS), models containing first generation 
leptoquarks can be constrained.

An upper limit $N_{lim}$ on the number of events coming from leptoquark 
induced processes can be obtained assuming Poisson distributions for the 
SM background expectation and for the signal.
For each contributing channel, we use the numbers of observed and 
expected events within a mass bin $\left[ M_{min}, M_{max} \right]$ 
of variable width, adapted to the expected mass resolution and
measured mass values for a given true LQ mass, and
which slides over the accessible mass range.
For example, only NC DIS (respectively CC DIS) candidates
$M_e \in \left[ 186 ; 204 \right] \GeV$ 
($M_h \in \left[ 170 ; 214 \right] \GeV$) will be used to constrain 
a $200 \GeV$ LQ undergoing a NC (CC) DIS-like decay.
For high LQ masses, the mass bin becomes very large because of the 
distortion of the mass spectrum mentioned in section~\ref{sec:pheno}. 
The evolution of $M_{min}$ as a function of the LQ mass is 
represented in Fig.~\ref{fig:cuts_inter}b for $ F = 0$ 
and $\mid F \mid = 2$ scalar LQs in the NC-like channel.
For LQs undergoing NC DIS-like decays,
typical signal detection efficiency including the optimized $y$ and mass
cuts is found to vary between $24 \%$
at $75 \GeV$, $\simeq 35 \%$ around $200 \GeV$ and $48 \%$
at $250 \GeV$.
In the CC DIS-like decay channel, this efficiency ranges between
$\simeq 19 \%$ for a $75 \GeV$ leptoquark, $\simeq 39 \%$ at
$100 \GeV$ and reaches $53 \%$ at $200 \GeV$.

In section~\ref{sec:brwlim}, we will consider the BRW model
where the only free parameter is the Yukawa coupling $\lambda$.
Only NC DIS-like data will be used to derive $N_{lim}$, which
can then be translated into an upper limit on
$\sigma_{LQ} + \sigma_{int}$, $\sigma_{LQ}$ being the part
of the $e^+ p \rightarrow e + q + X$ cross-section induced by
the LQ $s$- and $u$-channel processes, and $\sigma_{int}$
the interference term with DIS, both integrated  over the phase space
allowed by the mass dependent cuts applied in the analysis on
$M_e$ and $y_e$. 
The evolution of these integrated contributions $\sigma_{LQ}$ and $\sigma_{int}$
as a function of the LQ mass
is shown in Fig.~\ref{fig:cuts_inter}c for the $S_{0,R}$ ($\mid F \mid =2$) 
scalar leptoquark, as the dashed and dash-dotted lines respectively,
and in Fig.~\ref{fig:cuts_inter}d for the $S_{1/2,L}$ ($F=0$),
with emphasis on the high mass domain where the interference
contribution becomes visible.
In the former case the interference between LQ induced processes
and SM boson exchange is constructive, while it is destructive 
for the latter.
The cross-sections have here been calculated for a fixed value $\lambda=0.5$,
which is typical for the experimental sensitivity in the
displayed mass range. In Fig.~\ref{fig:cuts_inter}c the contribution 
$\sigma_u$ of
the $u$-channel $S_{0,R}$ exchange alone is also shown as the dotted line
(for the case depicted in Fig.~\ref{fig:cuts_inter}d, $\sigma_u$ is below
$10^{-4} \picob$ and is not represented in the figure).
As was mentioned in section 3, the mass dependent cuts applied on $M_e$ 
and $y_e$ considerably reduce the contributions of the interference and 
of virtual exchange (e.g. by a factor ${\cal{O}}(10)$ 
for a $S_{0,R}$ LQ at $M_{LQ} = 250 \GeV$).
As can be seen in Fig.~\ref{fig:cuts_inter}c and d,
the interference is negligible for $F=0$ leptoquarks and masses
up to $\simeq 275 \GeV$, but plays an important role for
$\mid F \mid =2$ LQs as soon as $M_{LQ} \simeq 220 \GeV$.
Moreover the $u$-channel contribution is always negligible.

In the mass domain where the interference between standard DIS and LQ induced 
processes can be neglected, we will move away from the BRW model and consider 
a more general case where the branching $\beta_e = \beta(LQ \rightarrow eq)$ 
is not determined by $\lambda$ only.
Taking $\beta_e$ and $\beta_{\nu} = \beta(LQ \rightarrow \nu q)$
as free parameters, NC and CC DIS-like data can be combined
to derive $N_{lim}$, which can then be translated into an upper
limit on the signal cross-section $ \sigma_{LQ} = \sigma_s + \sigma_u$ 
and thus on the Yukawa coupling $\lambda$. This will be
done in section~\ref{sec:liminccc}.
The signal cross-section in the mass domain considered being 
largely dominated by the
$s$-channel resonant production as mentioned above,
an upper limit on $\sigma_s \times \beta_e$ can also be derived.
Fixing $\lambda$, mass dependent upper limits on
the branching ratio $\beta_e$ can then be obtained, as will
be done in section~\ref{sec:betavsmass} using NC DIS data only.
In these two cases, we make use of the signal detection efficiencies
given above to translate $N_{lim}$ into an upper limit on the
signal cross-section.

The procedure which folds in, channel per channel, the
statistical and systematic errors is described in detail
in~\cite{H1LQ}.

\subsection{Mass Dependent Limits on the Yukawa Coupling in the BRW Model}
\label{sec:brwlim}

We first establish constraints on the BRW model described in
section~\ref{sec:pheno} taking into account all LQ induced contributions,
but restricting the analysis to NC DIS-like processes.
For the decay of resonantly produced LQs, the values of $\beta_e$ are specified and
given in table~\ref{tab:brwscalar}.
As discussed in section~\ref{sec:pheno}, the constraints can be extended 
beyond the kinematic limit by profiting from the tail expected in the 
$s$-channel towards low masses, and by properly taking into account the 
interference with SM boson induced processes.

\begin{figure}[tbh]
 \begin{center}
  \begin{tabular}{cc}
     \hspace*{-0.9cm}\mbox{\epsfxsize=0.55\textwidth
     \epsffile{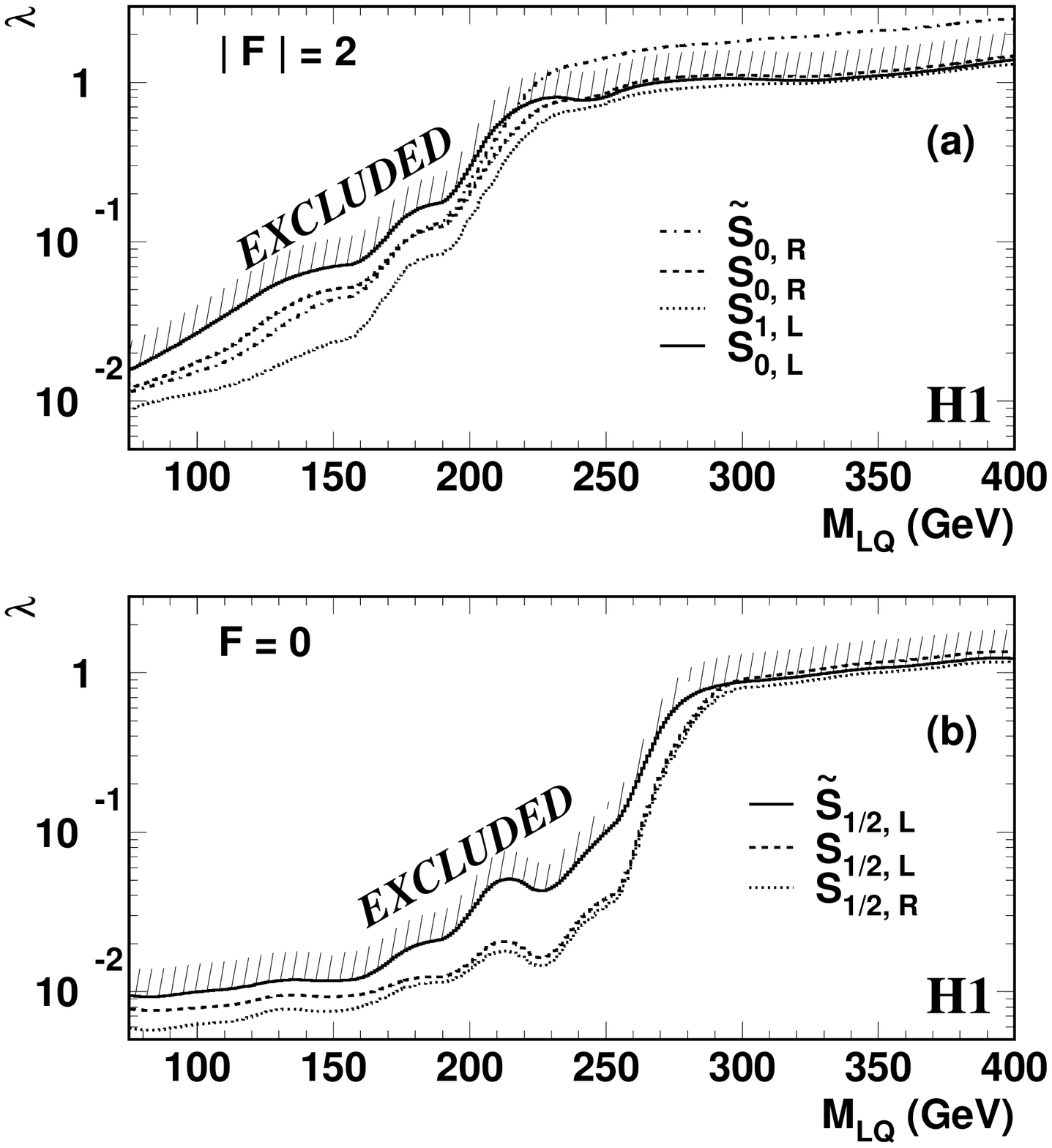}}
   &
     \hspace*{-0.8cm}\mbox{\epsfxsize=0.55\textwidth
      \epsffile{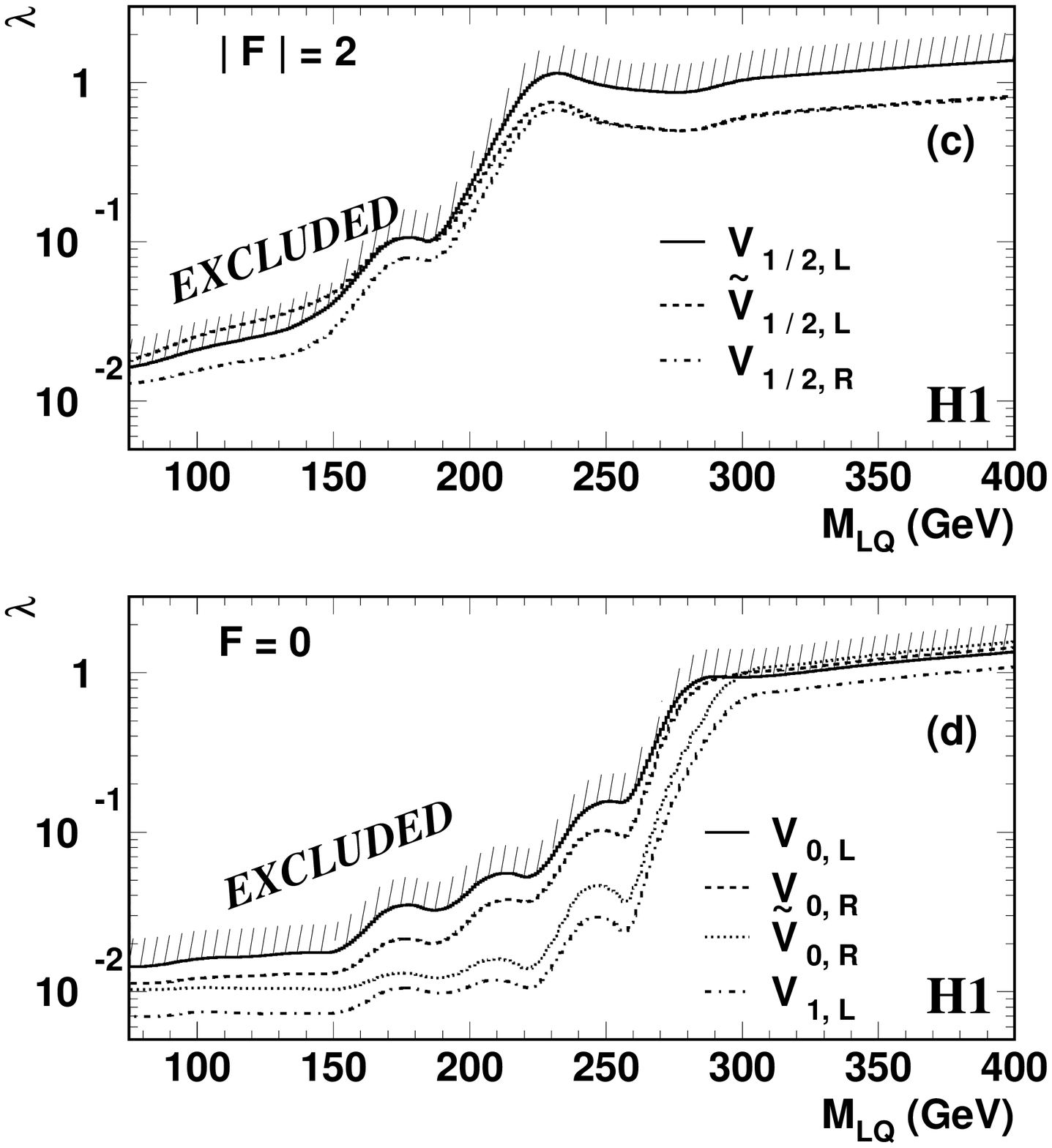}}
  \end{tabular}
 \caption[]{ \label{fig:limilambda}
 {\small Exclusion limits at $95 \%$ CL on the Yukawa coupling
         $\lambda$ as a function of the leptoquark mass for
         (a) $\mid F \mid = 2$ and (b) $F=0$ scalar, 
         (c) $\mid F \mid = 2$ and (d) $F=0$ vector 
         leptoquarks described by the BRW model.
         Domains above the curves are excluded. }}
 \end{center}
\end{figure}

In the very high mass domain, the interference $\sigma_{int}$ of LQ induced 
processes with NC DIS generally dominates the LQ cross-section $\sigma_{LQ}$.
Instead of $\sigma_{LQ}$ alone, we are thus directly sensitive 
to $\sigma_{sum}=\sigma_{LQ} + \sigma_{int}$, which, for those
LQ species which interfere destructively with DIS, could be negative
when integrated over the whole phase space, but remains positive
within the kinematic cuts applied.
We thus proceed the following way~:
\begin{itemize} 
 \item For a given LQ mass $M_{LQ}$, the numbers of observed and expected
       events within the optimized cuts $M_{min} < M_e < M_{max}$
       and $y_e > y_{cut}$ are used to set an upper limit $N_{lim}$
       on the number of signal events. 
       For LQ masses above the kinematic limit, the optimized $M_e$
       and $y_e$ cuts are nearly independent of $M_{LQ}$ and close to those
       displayed in Fig.~\ref{fig:cuts_inter}a,b for $M_{LQ} = 300 \GeV$.
       A first estimate $\lambda_0$ of the upper limit on $\lambda$ is then 
       obtained by solving :
       $N_{lim} = {\cal{L}} \sigma_{sum, cuts}(\lambda)$.
       $\sigma_{sum, cuts}$ is calculated analytically
       by integrating over the phase space allowed by the
       cuts the squared amplitude for the LQ process and its
       interference with NC DIS.
       The cuts applied, by reducing the NC DIS contribution,
       ensure that $\sigma_{sum, cuts}$ is positive for reasonable
       $\lambda$ values.
 \item The event generator LEGO is then used to produce LQ events
       at $(M_{LQ}, \lambda_0)$, from which we get the acceptance
       correction factor $A$ defined as the ratio of 
       the number of events satisfying the
       NC DIS selection cuts as well as the $M_e$ and $y_e$ cuts,
       to the number of events generated
       within these cuts. 
 \item To take into account next-to-leading order QCD corrections 
       on the LQ production cross-section, we
       calculate the convolution $K^*$ of the K-factor 
       given in~\cite{SPIRA}
       for the NWA, with the LQ Breit-Wigner distribution
       corresponding to $(M_{LQ}, \lambda_0)$.
       The resulting function $K^*(M_{LQ})$ will be used
       henceforth to calculate the
       signal cross-sections.
       The NLO corrections lead to a sizeable enhancement
       of the cross-section as will be seen below.
 \item The final rejected coupling is then obtained by solving 
       the equation
  $N_{lim} = {\cal{L}} \times A \times K^* \times \sigma_{sum, cuts}(\lambda)$.
\end{itemize}
The resulting rejection limits at $95 \%$ confidence level (CL)
are shown in Fig.~\ref{fig:limilambda} up to $M_{LQ} = 400 \GeV$
for $ \mid F \mid = 0$ or $2$ scalars or vectors. 
Constraints for even larger masses where one is probing
``strong'' (i.e. $\lambda >1$) coupling values in a contact
interaction will be discussed in
a separate paper.
The case of $ \mid F \mid = 0$ LQs, which can be produced via fusion
between the $e^+$ and an incident valence quark, naturally offers the
best sensitivity.
%

For LQs having $F=0$, these limits represent an improvement 
by a factor $\simeq 3$ compared to H1 published results~\cite{H1LQ}.
For a Yukawa coupling of the electromagnetic strength
$\lambda^2 / 4 \pi = \alpha_{EM}$ (i.e. $\lambda \simeq 0.3$),
such scalar (vector) LQs are excluded at $95 \%$ CL 
up to $275 \GeV$ ($284 \GeV$) by this analysis.
As can be seen in Fig.~\ref{fig:limilambda}b the $S_{1/2,R}$
is the scalar for which HERA has the highest sensitivity since
both charge states can be produced 
via a fusion $e^+ u$ or $e^+ d$. On the contrary, only an
$e^+ u$ ($e^+ d$) fusion is possible for the production 
of $S_{1/2,L}$ ($\tilde{S}_{1/2,L}$), 
for which the cross-sections are thus smaller. 
Note also that, due to the more favorable parton density,
a higher cross-section (and thus a better rejection limit)
is expected for $S_{1/2,L}$ than for $\tilde{S}_{1/2,L}$.
However at very high masses, the $S_{1/2,L}$ interferes destructively
with NC DIS, on the contrary to the $\tilde{S}_{1/2,L}$,
so that the limits become similar.

For $\mid F \mid = 2$ leptoquarks, the improvement compared to our previous
published results is less substantial since $e^-p$ data collected
in 1994 had been also taken into account in~\cite{H1LQ}.
These LQs will be best probed with $e^-p$ data which are being collected
since 1998.

For coupling values equal to the obtained limits,
NLO QCD corrections enhance the production cross-section
of a $F=0$ LQ by $\simeq 10 \%$ for $M_{LQ} = 100 \GeV$ to
$\simeq 30 \%$ at $250 \GeV$. For higher masses this enhancement
decreases because of the distortion of the LQ mass spectrum.
For $\mid F \mid = 2$ LQ, this enhancement 
remains below $\simeq 20 \%$.

$S_{1,L}$ and $S_{0,L}$ leptoquarks being allowed to undergo 
CC DIS-like decays with a branching ratio 
$\beta_{\nu} = 0.5$, 
the combination of NC and CC DIS analysis is expected to better
constrain these leptoquarks. 
The result of this combination is given for the $S_{0,L}$ by the 
curve $\beta_e = \beta_{\nu} = 0.5$ in Fig.~\ref{fig:limicc}a.
It can be seen that combining the two contributing channels improves 
the sensitivity on $S_{0,L}$ up to that achieved on $S_{0,R}$, recalled 
on Fig.~\ref{fig:limicc}a by the curve $\beta_e = 1$.

The DELPHI~\cite{DELPHI} experiment at LEP 
recently performed a direct search for single leptoquarks using data
accumulated at $e^+e^-$ centre of mass energies 
$\sqrt{s_{e^+e^-}}$ up to $183 \GeV$.
Constraints relevant for some LQ species have also been obtained
indirectly by ALEPH~\cite{ALEPH}, OPAL~\cite{OPAL}
and L3~\cite{L3} Collaborations from measurements 
of hadronic cross-sections and asymmetries at
$\sqrt{s_{e^+e^-}} = 130$ to 183 GeV.
For a Yukawa coupling of the electromagnetic strength, 
the direct search via single production results in a limit
of $171 \GeV$ which is not yet competitive with this H1 analysis.
For similar coupling values, the indirect search at LEP is yet only
competitive for the $S_{0,L}$, $S_{1,L}$, $V_{0,L}$, $\tilde{V}_{0,R}$
and $V_{1,L}$, for which limits up to $240 \GeV$ and $470 \GeV$
have been obtained in the scalar and vector case respectively.
However the sensitivity of the indirect search drops
quickly with the Yukawa coupling, and for $\lambda=0.1$
bounds lie below $100 \GeV$ for all LQ species~\cite{OPAL}.


In constrast to an $ep$ collider, the limits derived at the TeVatron
where LQs are mostly produced in pair via the strong interaction,
with larger cross-sections expected for vector than for
scalar leptoquarks,
are essentially independent of the Yukawa coupling.
Recent results have been published for scalar leptoquarks by
$D\emptyset$~\cite{D01GENE} and CDF~\cite{CDF1GENE} Collaborations. 
In the BRW model, a combination of these 
results~\cite{TEVCOMBINED} excludes scalar leptoquark
masses up to $242 \GeV$.
Comparisons of our results with limits obtained at the TeVatron
collider in more general models will be presented in
sections~\ref{sec:liminccc} and~\ref{sec:betavsmass}.
%
%
%

\subsection{Mass Dependent Limits on the Yukawa Coupling in Generic Models}
\label{sec:liminccc}

\begin{figure}[tbh]
 \begin{center}
  \begin{tabular}{cc}
     \hspace*{-0.9cm}\mbox{\epsfxsize=0.50\textwidth
     \epsffile{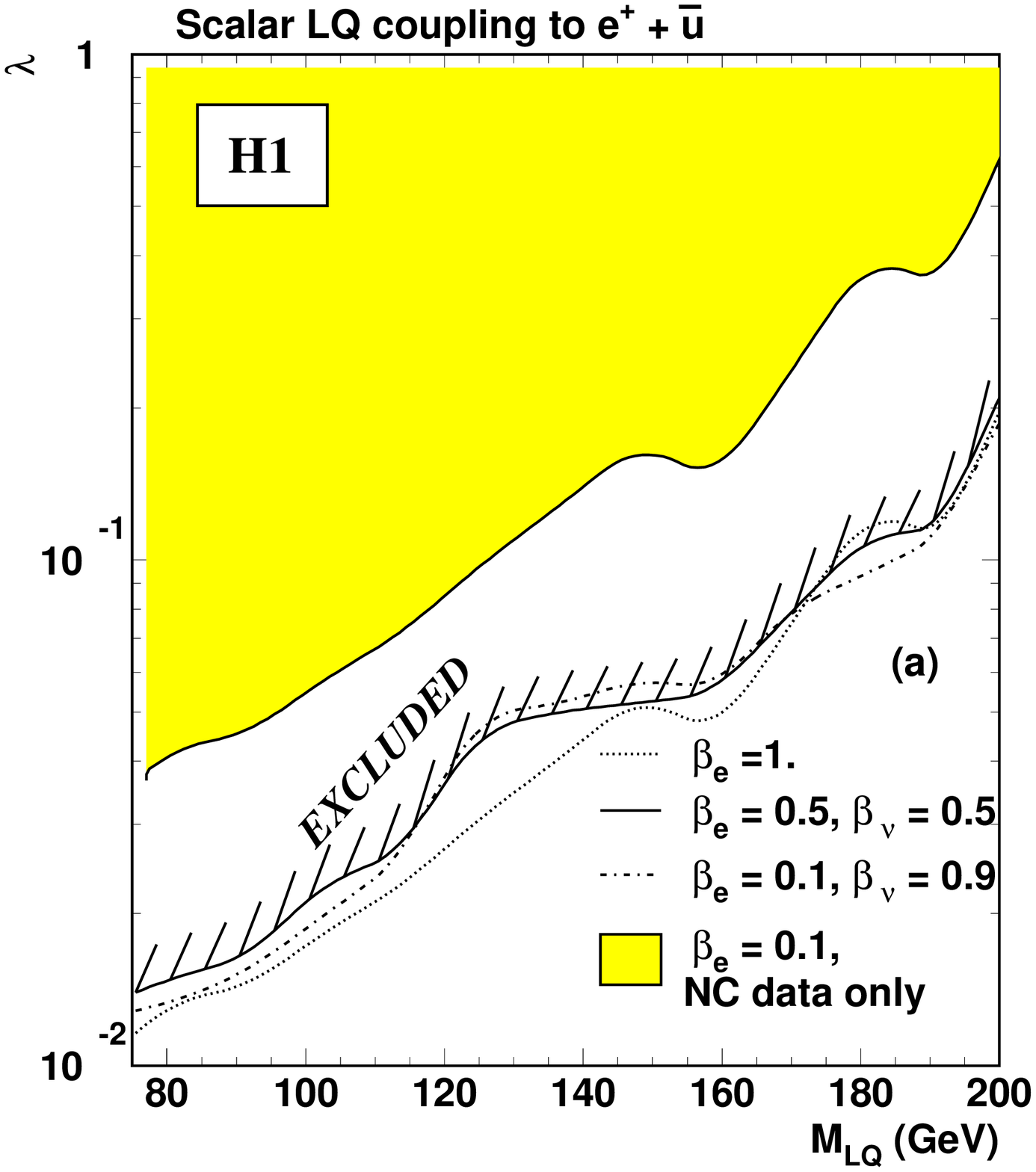}}
   &
     \hspace*{-0.8cm}\mbox{\epsfxsize=0.50\textwidth
     \epsffile{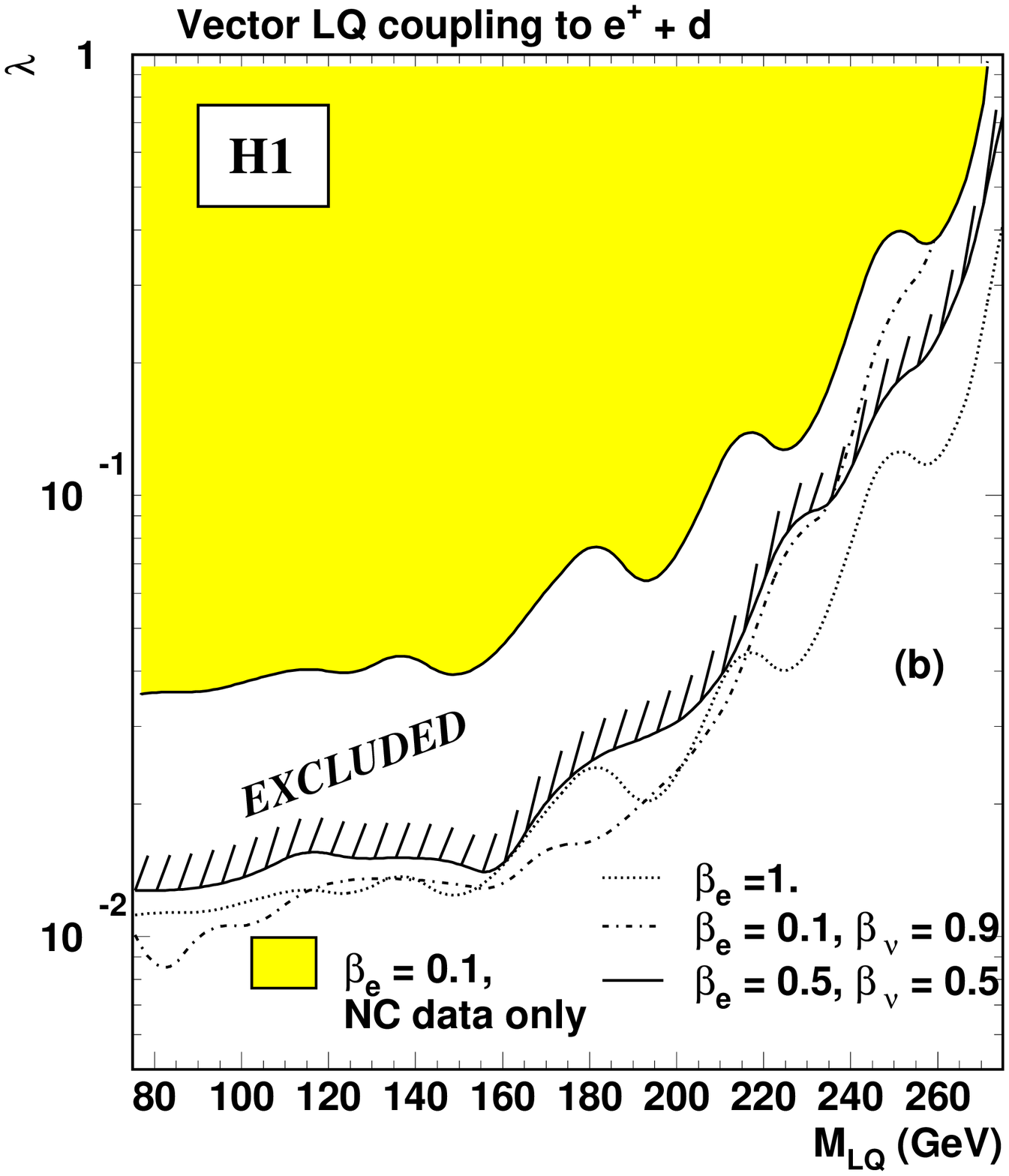}}
  \end{tabular}
 \caption[]{ \label{fig:limicc}
 {\small Exclusion limits at $95 \%$ CL on the Yukawa coupling
         $\lambda$ as a function of the leptoquark mass for
         (a) scalar leptoquarks produced via $e^+ \bar{u}$ fusion
         and (b) vector leptoquarks produced via $e^+ d$ fusion.
         Different hypotheses for the branching ratios into
         $eq$, $\nu q$ are considered.
         Domains above the curves are excluded. }}
 \end{center}
\end{figure}

Moving away from the BRW model, we now consider leptoquarks
for which the branching ratios $\beta_e$ and $\beta_{\nu}$
in NC and CC DIS-like decays are free parameters.
As an example, the $95 \%$ CL exclusion contour for a scalar LQ
with $\mid F \mid = 2$ decaying with $\beta_{\nu} = 90 \%$ in $\nu q$ and 
$\beta_e = 10 \%$ in $eq$ is shown in Fig.~\ref{fig:limicc}a as the 
dash-dotted line.
In this case, limits are shown only in the mass domain where the interference
of LQ processes with DIS can be neglected.
The gain obtained by the combination of both channels can be seen
when comparing this contour with the greyed
domain showing the sensitivity achieved using only NC DIS-like channel.
For $\beta_{\nu} = 90 \%$ and $\beta_e = 10 \%$, scalar LQ masses
below $200 \GeV$ are excluded at $95 \%$ CL by our analysis,
for Yukawa couplings of the electromagnetic strength.
This extends far beyond the domain excluded by TeVatron 
experiments~\cite{D01GENE,CDF1GENE}
which for such small values of $\beta_{e}$ only exclude scalar leptoquark
masses below $\simeq 110 \GeV$.
At HERA the sensitivity will be considerably enhanced with an electron incident
beam where these LQs can be produced via a fusion with a {\it{valence}} 
$u$ quark. 
Hence a large discovery potential
is opened for HERA, for high mass leptoquarks decaying in $\nu q$
with a high branching ratio.
%
%

Similar results are given for a vector LQ coupling to $e^+ d$
in Fig.~\ref{fig:limicc}b.
For the above values of ($\beta_e, \beta_{\nu}$) and of the
Yukawa coupling, the excluded mass
domain extends in this case to $250 \GeV$.

\subsection{Mass Dependent Limits on 
            {\boldmath $\beta(LQ \rightarrow eq)$ }}
\label{sec:betavsmass}
 
\begin{figure}[tbh]
 \begin{center}
  \epsfxsize=0.75\textwidth
  \epsffile{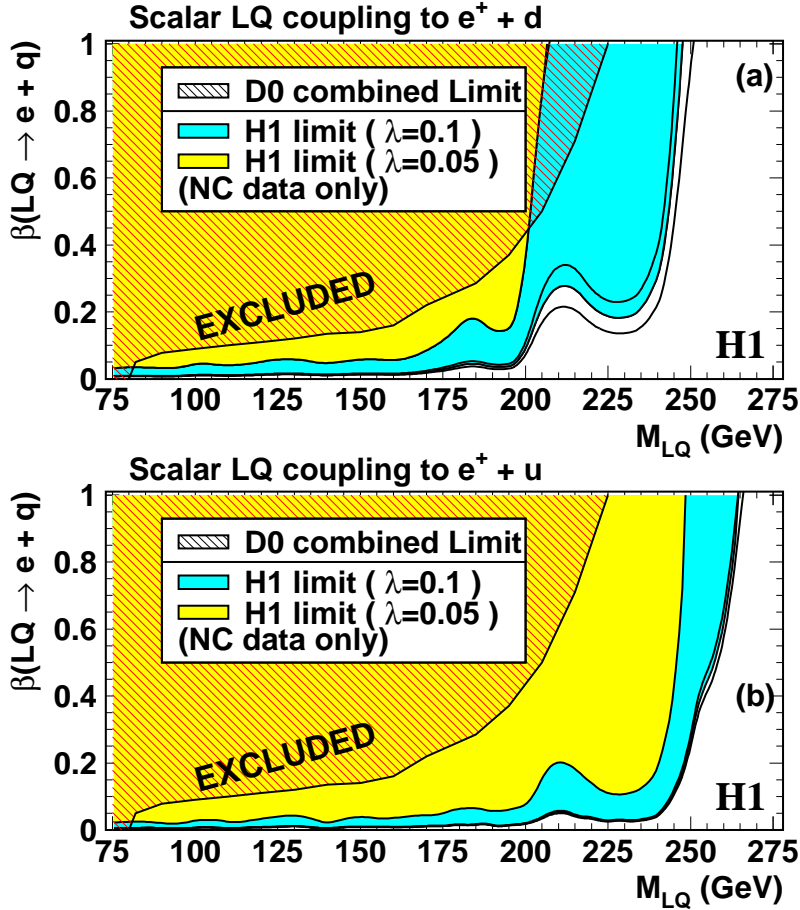}
 \caption[]{ \label{fig:limibeta}
 {\small Mass dependent exclusion limits at $95 \%$ CL on
         the branching ratio $\beta(LQ \rightarrow eq)$
         for scalar leptoquarks produced by (a) $e^+ d$ and
         (b) $e^+ u$ fusion. 
         Two exclusion regions corresponding to $\lambda=0.1$ and
         $\lambda = 0.05$ are represented as greyed areas.
         For $\lambda = 0.1$, the error bands on the exclusion 
         limits in (a) and (b) 
         illustrate the sensitivity to $d$ and $u$ quark densities
         respectively (see text).
         The $D\emptyset$ limit is also shown as hatched region. }}
 \end{center}
\end{figure}

We consider here leptoquarks which undergo NC DIS-like decays
with a branching ratio $\beta_e$ and do not make any
assumption on the other possible decay modes of the LQ.
For a fixed value of the Yukawa coupling $\lambda$
upper limits on LQ production cross-section are translated
in terms of mass dependent limits on the branching $\beta_e$.
Exclusion limits at $95 \%$ CL are shown in Fig.~\ref{fig:limibeta}a,b 
as greyed areas for scalar LQs produced via a $e^+d$ or $e^+u$ fusion 
respectively.
%
%
On Fig.~\ref{fig:limibeta}a, the central limit contour for
$\lambda = 0.1$ has been obtained assuming an uncertainty
on the $d$ quark density distribution which varies with
$x$ as mentioned in section~\ref{sec:pheno}.
Curves above and below this contour define the error band
obtained when the lack of knowledge on the proton structure
is directly applied on the parton densities, instead of entering
as a systematic uncertainty~: the $d$ quark density is enhanced/lowered
by a factor varying linearly between $7\%$ at low masses and
up to $ 30 \%$ at $250 \GeV$.
For LQs coupling to $e^+ u$ (Fig.~\ref{fig:limibeta}b), a constant scaling 
factor of $\pm 7\%$ has been applied on the quite well-known $u$ density, 
resulting in a much narrower error band. 

Despite the small $\lambda$ values considered,
domains excluded by this analysis extend
beyond the region covered by the $D\emptyset$ experiment~\cite{D01GENE}
at the TeVatron also in the less favourable case of LQs coupling to $e^+ d$.
This is especially the case for small values of $\beta_e$.
For example, for $\beta_e = 10 \%$, this analysis rules out masses below 
$\simeq 240 \GeV$ ($\simeq 200 \GeV$) if
$\lambda=0.1$ ($\lambda=0.05$) independently
of other possible decay modes of the leptoquark, as can be seen in
Fig.~\ref{fig:limibeta}b. This limit extends up to
$260 \GeV$ for an electromagnetic strength of the Yukawa
coupling $\lambda$ as will be shown by the greyed domain in
Fig.~\ref{fig:limibetatau}a.
Hence, a yet unexplored region in the high mass - low $\beta_e$
domain is probed by this analysis.

\section{Constraints on Couplings to Mixed Generations}
\label{sec:lqmix} 

In this section, we consider LFV LQs which couple
to both the electron and a second or third generation lepton.

The case of low mass ($M < \sqrt{s_{ep}}$) LFV LQs is first
addressed in section~\ref{sec:lfvlow}.
The analysis is there 
restricted to LQs possessing a coupling $\lambda_{11}$, allowing
the LQ to be produced or exchanged between the incident lepton
and a {\it{valence}} quark coming from the proton.
Moreover we do not consider $e \leftrightarrow \mu$ transitions
induced by a low mass LQ, such processes being strongly
constrained by low energy experiments as will be seen in
section~\ref{sec:lfvhigh} where high mass
($M > \sqrt{s_{ep}}$) LFV LQs are considered. In this latter
case, the study will be extended to any $\lambda_{1i}$,
and both $\lambda_{2j}$ and $\lambda_{3j}$.

\subsection{Low Mass {\boldmath $(M < \sqrt{s_{ep}})$} LFV Leptoquarks}
\label{sec:lfvlow}

%
\begin{figure}[hbt]
 \begin{center}
  \begin{tabular}{cc}
     \hspace*{-0.9cm}\mbox{\epsfxsize=0.50\textwidth
     \epsffile{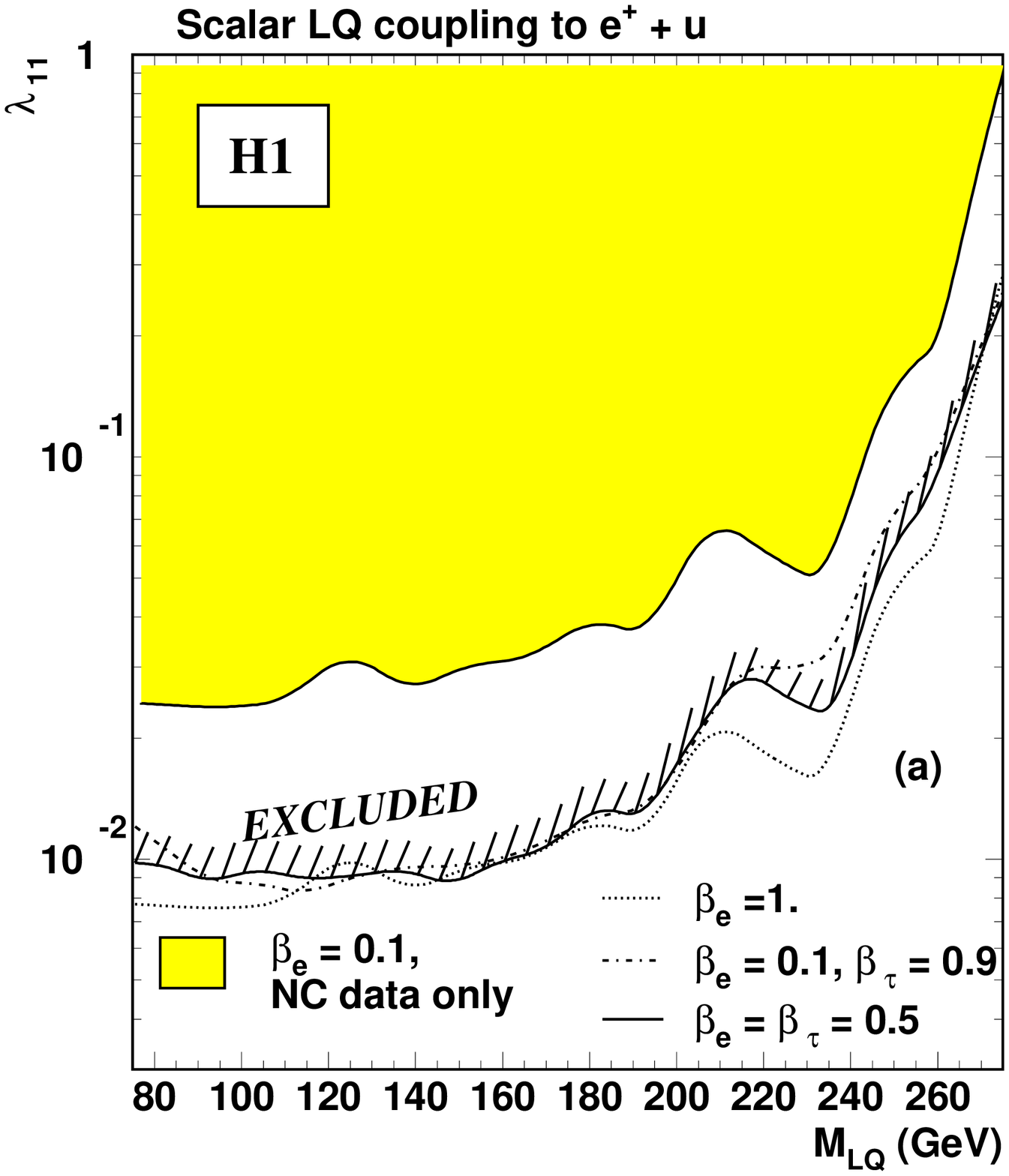}}
   &
     \hspace*{-0.8cm}\mbox{\epsfxsize=0.50\textwidth
     \epsffile{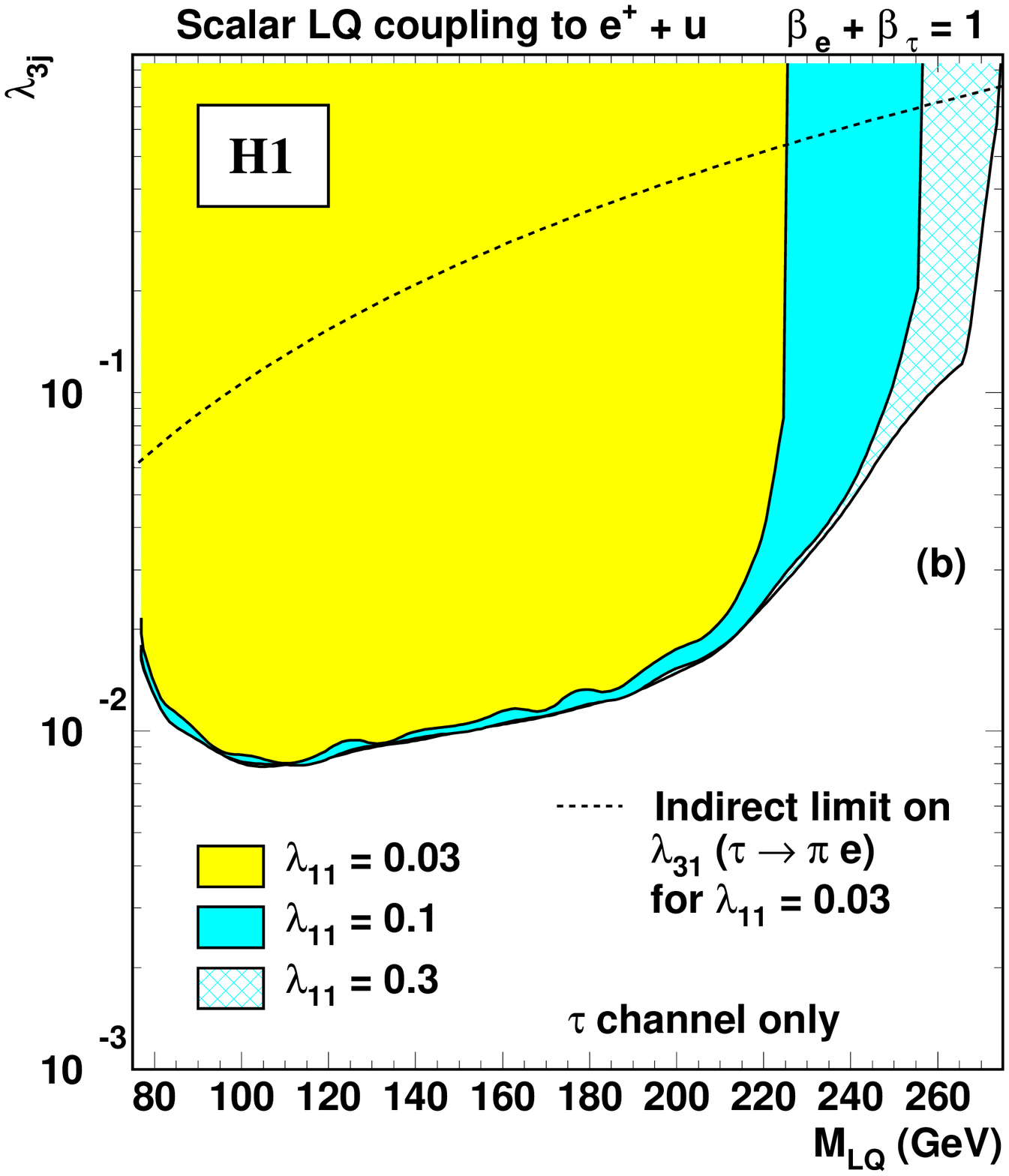}}
  \end{tabular}
 \caption[]{ \label{fig:limibetatau}
 {\small (a) Mass dependent exclusion limits at $95 \%$ CL
         on the Yukawa coupling $\lambda_{11}$, for scalar
         leptoquarks produced by $e^+ u$ fusion. 
         Different hypotheses for the branching ratios into
         $eq$, $\tau q$ are considered.
         Domains above the curves are excluded. 
         (b) Exclusion domains in the plane $\lambda_{3j}$
         $(j=1,2)$ against the leptoquark mass for several
         fixed values of $\lambda_{11}$ (greyed areas).
         A mass dependent indirect limit on $\lambda_{31}$ is
         also represented by the dashed line. }}
 \end{center}
\end{figure}
We now consider LFV LQs possessing a coupling $\lambda_{11}$
to first generation lepton-quark pairs as well as a coupling
$\lambda_{3j}$ with leptons of the third generation.

Mass dependent exclusion limits on $\lambda_{11}$ at $95\%$ CL are shown in 
Fig.~\ref{fig:limibetatau}a when fixing the relative branching fractions 
$\beta_e$ and $\beta_{\tau}$ into $e + jet$ and $\tau + jet$ final states.
A generic scalar leptoquark coupling to $e^+ + u$ pairs (such as the 
$S_{1/2,L}$ in the BRW model) has been considered for three different 
sets of $(\beta_e,\beta_{\tau})$.
Here both $e^+ + jet$ and $\tau^+ + jet$ channels are combined.
This latter channel is background free but the former benefits from a higher 
selection efficiency,
such that finally both provide a comparable sensitivity to the signal.
Thus, as soon as $\beta_e + \beta_{\tau}$ approaches $1$, the limits 
derived are very similar to those obtained for $\beta_e=1$.
Assuming $\beta_e = 10 \%$ and $\beta_{\tau} = 90 \%$, masses below 
$275 \GeV$ are excluded at $95 \%$ CL for an electromagnetic strength of 
the $\lambda_{11}$ coupling.
Such limits are especially interesting since, for small $\beta_e$ and 
high $\beta_{\tau}$, the mass domain excluded by the TeVatron 
does not extend very far.
The CDF experiment has performed a search for third generation LQs
looking at $\tau \tau b b $ final states~\cite{CDF3GENE}, and excluded scalar
LQs with masses below $99 \GeV$ if $\beta (LQ \rightarrow \tau b) = 1$.
A complementary search has been carried out by $D\emptyset$~\cite{D03GENE},
where the analysis of $\nu \nu b b $ final states leads to a
lower mass limit of $94 \GeV$ for $\beta (LQ \rightarrow \nu b) = 1$.
HERA thus appears to provide access to an unexplored domain for
LQs decaying with a small branching ratio in $eq$ and a high
branching ratio in $\tau q$.

An alternative representation of our results is given in 
Fig.~\ref{fig:limibetatau}b in the plane $\lambda_{3j}$ against 
the LQ mass, for different fixed values of $\lambda_{11}$.
We consider here a scalar LQ formed via $e^+ u$ fusion (carrying 
the quantum numbers of the $S_{1/2,L}$ in the BRW model)
such that only couplings $\lambda_{3j}$ with $j=1,2$ are relevant
and make the additional assumption that $\beta_e + \beta_{\tau} =1$.
The $\tau$ + jet final states analysis provides a sensitivity on
$\lambda_{31}$ so long as the LQ is light enough to have
a substantial production cross-section via $\lambda_{11}$, as
can be seen on the $95 \%$ CL exclusion domains shown
in Fig.~\ref{fig:limibetatau}b.
For both couplings $\lambda_{11}$ and $\lambda_{3j}$ of the electromagnetic
strength, such LQs lighter than $270 \GeV$ are excluded at
$95 \%$ CL. 
This limit still reaches $237 \GeV$ for LQs formed by $e^+ d$ fusion 
(not shown).
A similar lepton flavour violation analysis has been carried out by ZEUS
Collaboration~\cite{ZEUSLFV} using an integrated luminosity of 
$\simeq 3 \picob^{-1}$. 
For $\lambda_{11}=\lambda_{3j}=0.03$, the analysis presented here extends 
their excluded mass range by $\simeq 65 \GeV$.
Also shown in Fig.~\ref{fig:limibetatau}b is the best indirect 
limit~\cite{DAVIDSON} on $\lambda_{31}$ in the case $\lambda_{11}=0.03$. 
This indirect constraint comes from the upper limit on the branching ratio
$\beta( \tau \rightarrow \pi^0 e)$ which could be affected by the process 
$\tau^+ \rightarrow \bar{u} + LQ$ followed by the $e^+ + u$ decay of the 
virtual leptoquark.
Here the most recent upper limit~\cite{PDG98} on 
$\beta( \tau \rightarrow \pi^0 e)$ has been used to update the bound derived 
in~\cite{DAVIDSON}.
The H1 direct limit improves this indirect constraint by typically
one order of magnitude. 
No low energy process constrains the coupling products involving 
$\lambda_{32}$. 
In this case, H1 covers a yet unexplored domain.

It should be noted however that for most other LFV LQ species (except 
namely $\tilde{V}_{0,R}$, $S_{0,R}$ and $\tilde{V}_{1/2,L}$)
the coupling products involving $\lambda_{32}$ and $\lambda_{33}$
can be constrained by low energy experiments, in
particular by $\tau \rightarrow K^0 e$,
$B \rightarrow \tau e X$, $V_{ub}$ measurements or 
$K \rightarrow \pi \nu \bar{\nu}$~\cite{DAVIDSON}.
This latter process yields the most severe bound, relevant for
$\lambda_{11} \times \lambda_{32}$, but which for $F=0$ LQs
only applies to those possessing the quantum numbers
of the $V_{1,L}$ in the BRW model.
It has been checked (not shown here) that for all other $F=0$ LQs, 
the H1 direct limits on these coupling products 
extend beyond the domain covered by low energy phenomena. 
More detailed comparisons of HERA sensitivity with
indirect bounds will be discussed in the next section.

\subsection{High Mass {\boldmath $(M > \sqrt{s_{ep}})$} LFV Leptoquarks}
\label{sec:lfvhigh}

In this section, we make use of the fact that no $\mu + jet$ or 
$\tau + jet$ candidate was found 
(with kinematic properties compatible with a
$2 \rightarrow 2$ body process) to set constraints on very high mass 
LFV leptoquarks.
For LQ masses well above the kinematic limit, the cross-section 
$\sigma (e^+ + p \rightarrow l_n^+ + jet + X)$ depends only
on $( \lambda_{1i} \lambda_{nj} / M^2_{LQ} )^2$, with $i$ and $j$ 
indexing the generation of the quark coupling to $LQ-e$ and $LQ-l_n$ 
respectively, and where $l_n = \mu$ for $n=2$ and $l_n = \tau$ for $n=3$.

\begin{figure}[tbh]
 \begin{center}
  \epsfxsize=0.8\textwidth
    \epsffile{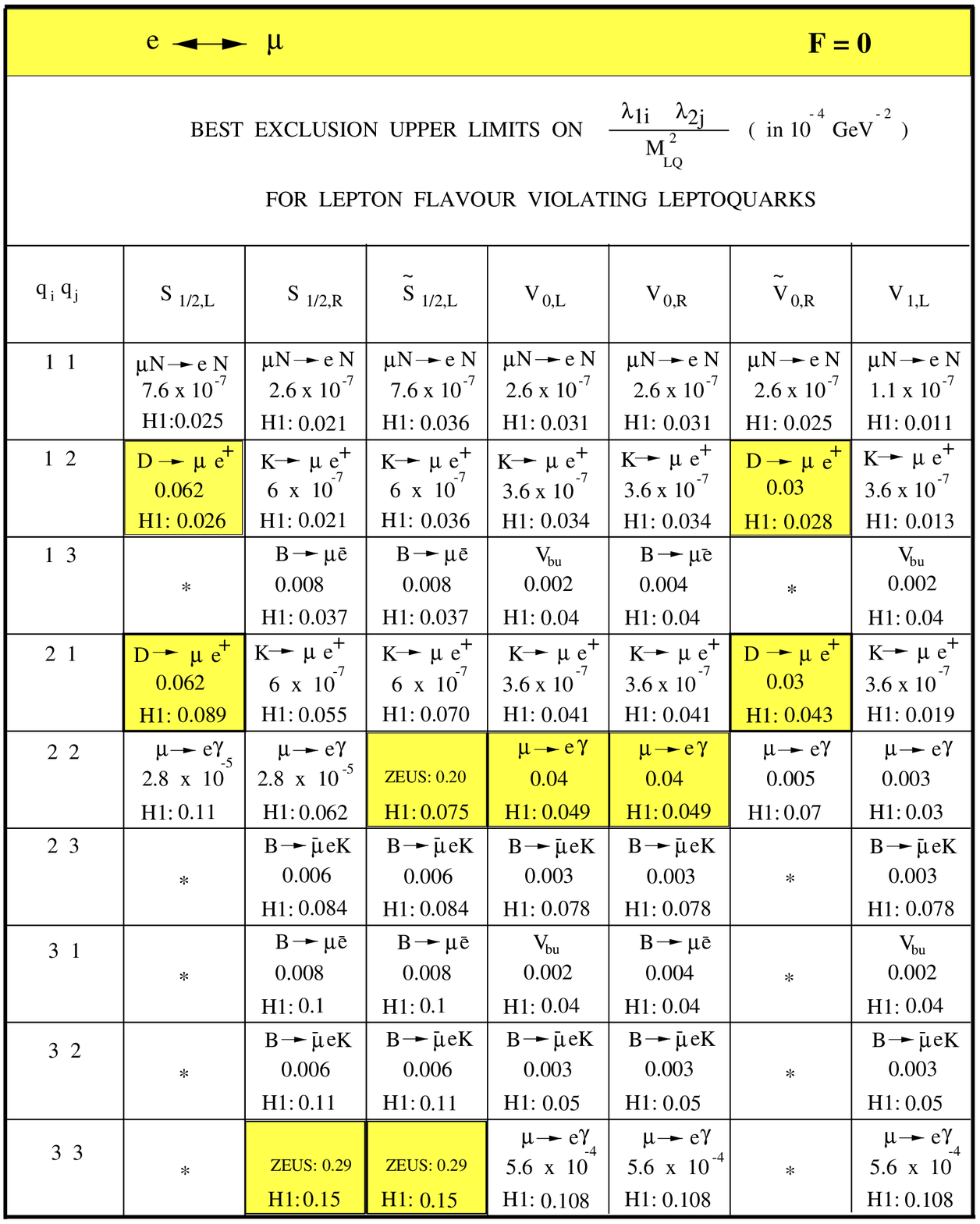}
 \caption[]{ \label{fig:lfvmu0}
 {\small Rejection limits at $95\%$ on
         $\lambda_{1i} \lambda_{2j} / M^2_{LQ}$ in units
         of $10^{-4} \GeV^{-2}$ for $F=0$ leptoquarks, compared
         to constraints from indirect processes. 
         The first column indicates the generations of the
         quarks $q_i$ and $q_j$ coupling respectively
         to $LQ - e$ and $LQ - \mu$. 
         In each box, the process which provides a 
         most~\cite{DAVIDSON} stringent indirect constraint 
         is listed (first line) together with its exclusion limit
         (second line) and compared to the actual H1 result (third
         line). 
         Shadowed boxes emphasize where HERA limit is 
         comparable to (within a factor of 2) or better than
         the indirect constraints.
         The open boxes marked with a $*$ are cases 
         which would involve a top quark. }}  
 \end{center}
\end{figure}

\begin{figure}[tbh]
 \begin{center}
  \epsfxsize=0.8\textwidth
    \epsffile{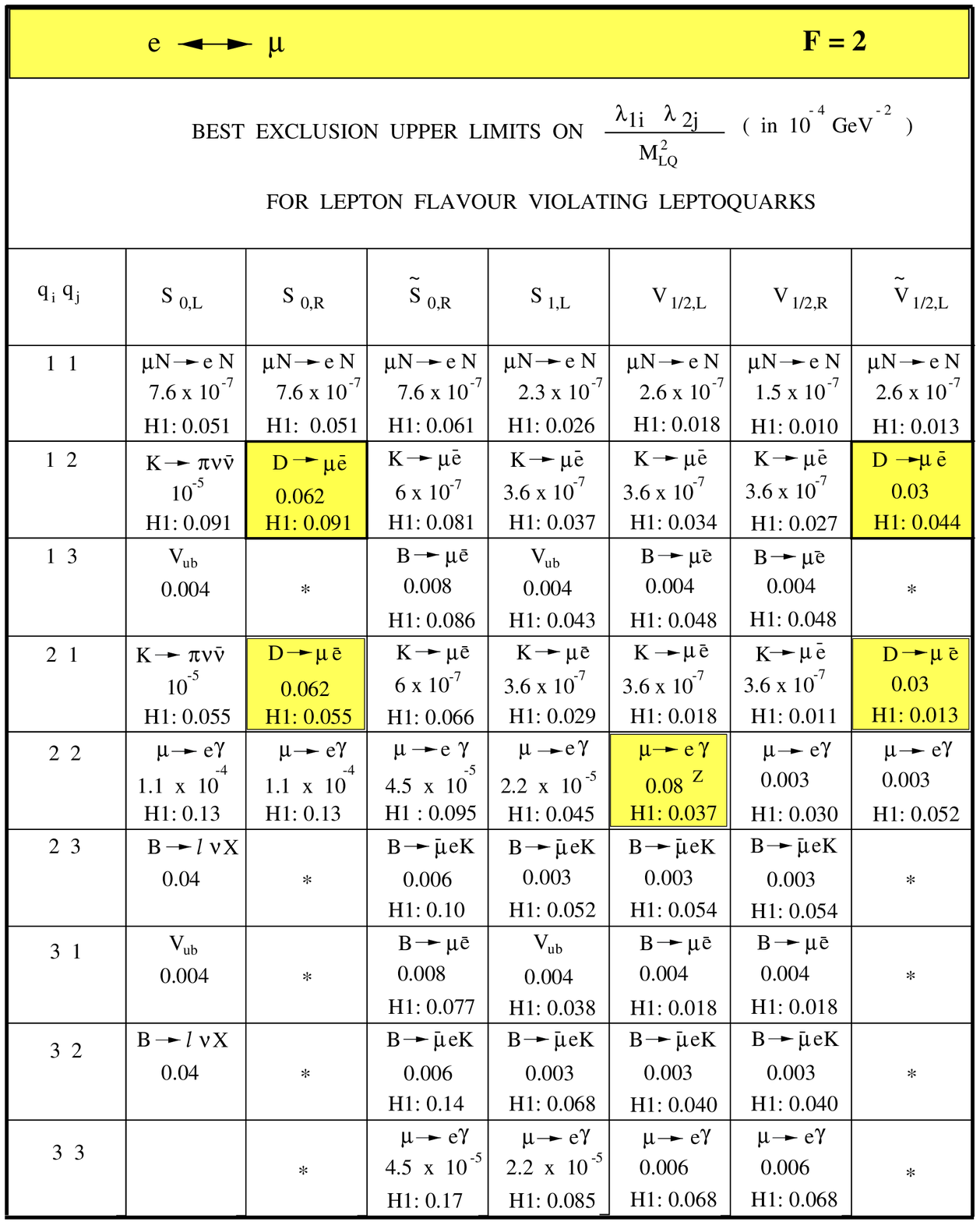}
 \caption[]{ \label{fig:lfvmu2}
 {\small Rejection limits on
         $\lambda_{1i} \lambda_{2j} / M^2_{LQ}$ in units
         of $10^{-4} \GeV^{-2}$ for $\mid F \mid = 2$ leptoquarks,
         compared to constraints from indirect processes.
         The first column indicates the generations of the
         quarks $q_i$ and $q_j$ coupling respectively
         to $LQ - e$ and $LQ - \mu$.
         In each box, the process which provides a 
         most~\cite{DAVIDSON} stringent indirect constraint 
         is listed (first line) together with its exclusion limit
         (second line) and compared to the actual H1 result (third
         line). 
         Shadowed boxes emphasize where HERA limit is 
         comparable to (within a factor of 2) or better than
         the indirect constraints.
         The superscripts $Z$ indicate where earlier
         HERA~\cite{ZEUSLFV} results already improved indirect bounds.
         The open boxes marked with a $*$ are cases 
         which would involve a top quark. }}
 \end{center}
\end{figure}

\begin{figure}[tbh]
 \begin{center}
  \epsfxsize=0.8\textwidth
    \epsffile{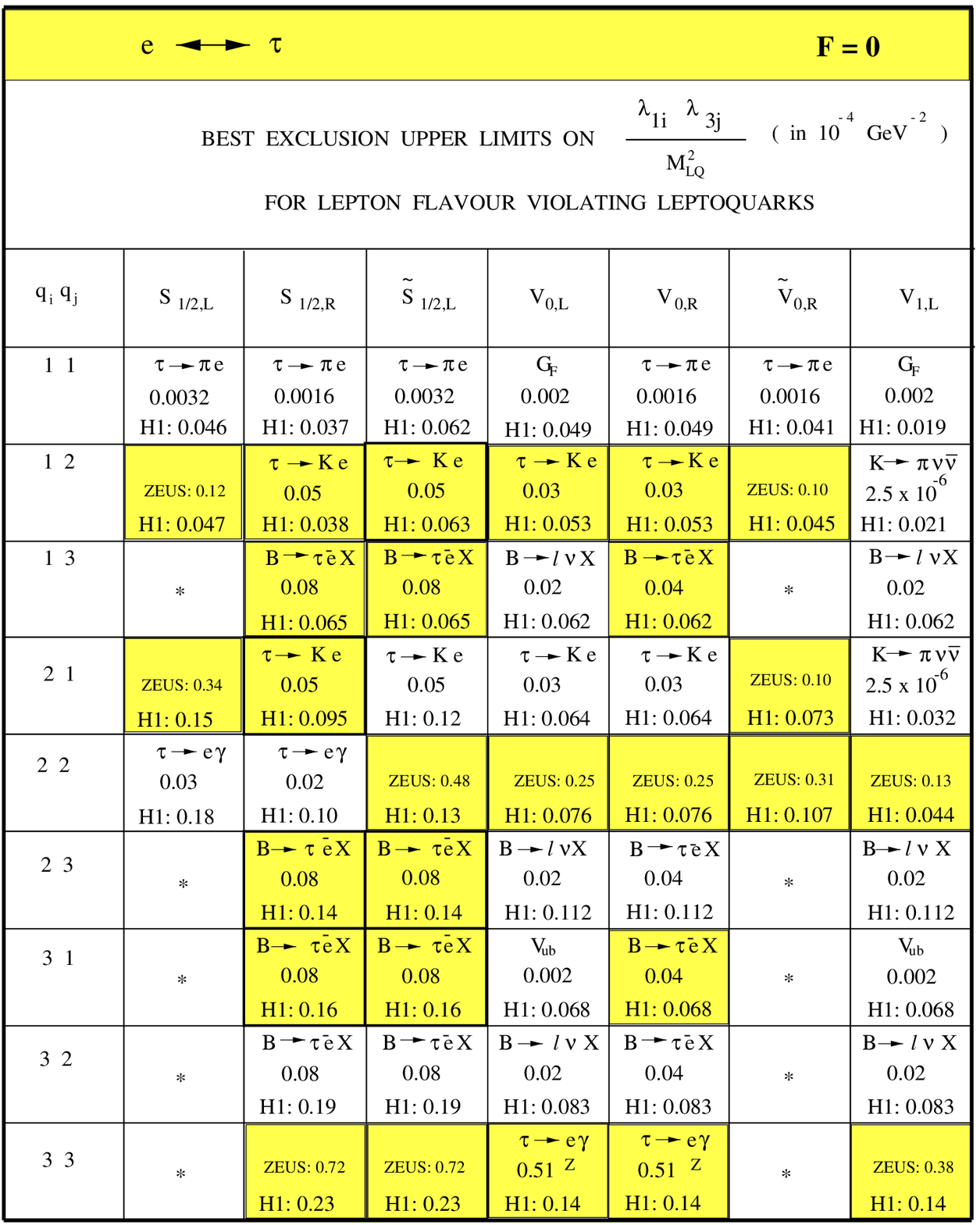}
 \caption[]{ \label{fig:lfvtau0}
 {\small Rejection limits on
         $\lambda_{1i} \lambda_{3j} / M^2_{LQ}$ in units
         of $10^{-4} \GeV^{-2}$ for $F=0$ leptoquarks,
         compared to constraints from indirect processes.
         The first column indicates the generations of the
         quarks $q_i$ and $q_j$ coupling respectively
         to $LQ - e$ and $LQ - \tau$.
         In each box, the process which provides a 
         most~\cite{DAVIDSON} stringent indirect constraint 
         is listed (first line) together with its exclusion limit
         (second line) and compared to the actual H1 result (third
         line). 
         For the $\tilde{S}_{1/2,L}$, which does not
         couple to the neutrino, limits on
         $\lambda_{11} \times \lambda_{32}$ and on
         $\lambda_{12} \times \lambda_{31}$ derived in~\cite{DAVIDSON}
         from $K \rightarrow \pi \nu \bar{\nu}$ have been
         replaced by the bounds obtained from $\tau \rightarrow K e$.
         Shadowed boxes emphasize where HERA limit is 
         comparable to (within a factor of 2) or better than
         the indirect constraints.
         The superscripts $Z$ indicate where earlier
         HERA~\cite{ZEUSLFV} results already improved indirect bounds.
         The open boxes marked with a $*$ are cases 
         which would involve a top quark. }}     
 \end{center}
\end{figure}

\begin{figure}[tbh]
 \begin{center}
  \epsfxsize=0.8\textwidth
    \epsffile{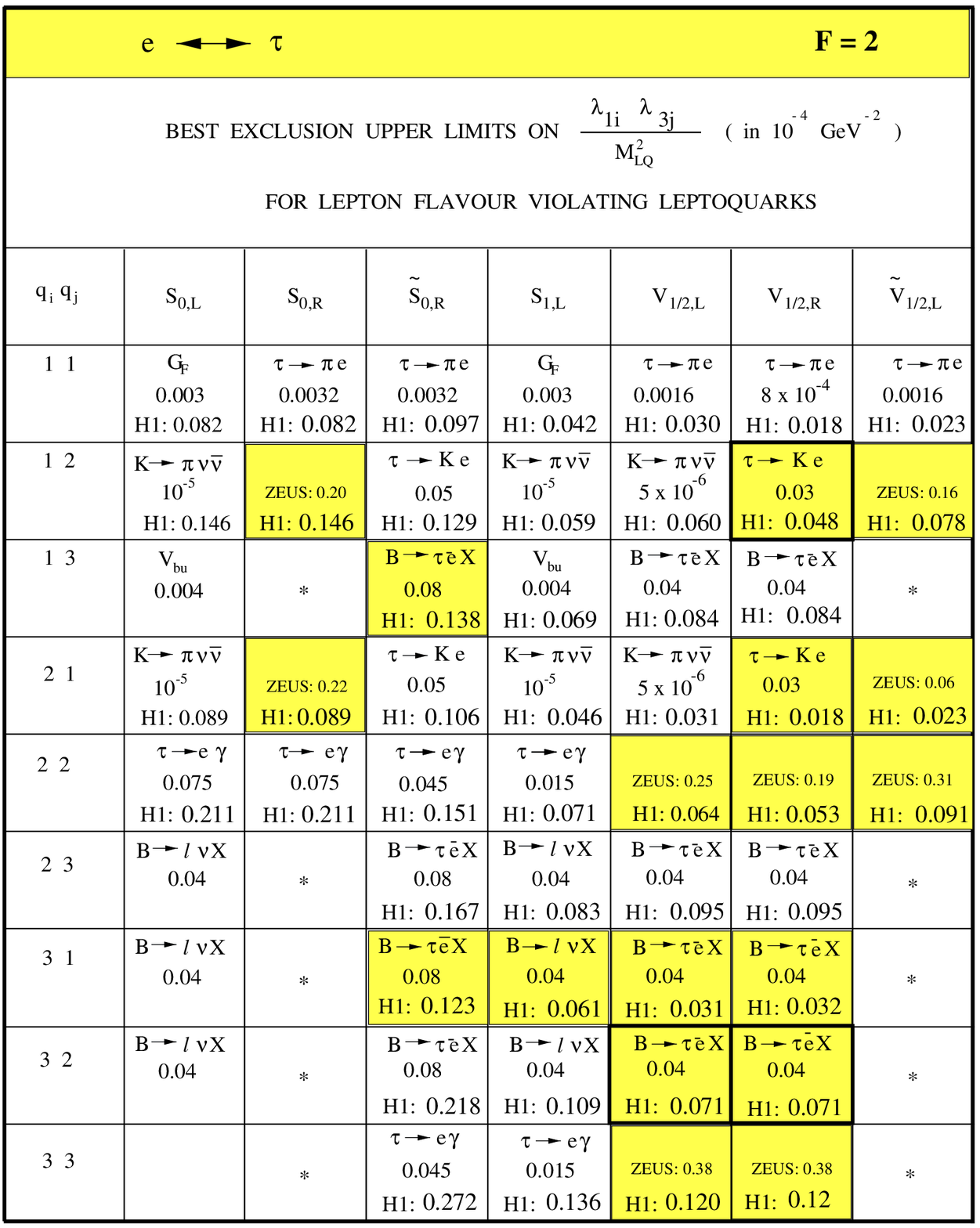}
 \caption[]{ \label{fig:lfvtau2}
 {\small Rejection limits on
         $\lambda_{1i} \lambda_{3j} / M^2_{LQ}$ in units
         of $10^{-4} \GeV^{-2}$ for $\mid F \mid = 2$ leptoquarks,
         compared to constraints from indirect processes.
         The first column indicates the generations of the
         quarks $q_i$ and $q_j$ coupling respectively
         to $LQ - e$ and $LQ - \tau$.
         In each box, the process which provides a 
         most~\cite{DAVIDSON} stringent indirect constraint 
         is listed (first line) together with its exclusion limit
         (second line) and compared to the actual H1 result (third
         line). 
         Shadowed boxes emphasize where HERA limit is 
         comparable to (within a factor of 2) or better than
         the indirect constraints.
         The open boxes marked with a $*$ are cases 
         which would involve a top quark. }}
 \end{center}
\end{figure}

The $95 \%$ CL rejection limits are given for a $e \leftrightarrow \mu$ 
transition in Fig.~\ref{fig:lfvmu0} for $F=0$ and Fig.~\ref{fig:lfvmu2}
for $\mid F \mid = 2$ leptoquarks.
Results are obtained for all possible quarks involved, $q_i$ being the 
quark coupling to $LQ - e$ and $q_j$ the one coupling to $LQ - \mu$. 
The limits are given in units of $10^{-4} \GeV^{-2}$.
For processes involving a $b$ quark in the initial state ($i=3$ or $j=3$),
it has been checked that the correction to the cross-section due to the
finite mass of the $b$ remains below $\sim 5 \%$~\cite{SPIRAEB}.
Early HERA results were presented in a similar representation 
by the ZEUS Collaboration in~\cite{ZEUSLFV}.
The nomenclature defined in table~\ref{tab:brwscalar}
has been kept to distinguish all possible LQ quantum numbers.
Also given for each entry in the depicted tables are the constraints from 
the indirect process which currently provides the most stringent 
bound~\cite{DAVIDSON}.
Figures~\ref{fig:lfvtau0} and~\ref{fig:lfvtau2} show similar
limits for the $e \leftrightarrow \tau$ transition.

In Figs.~\ref{fig:lfvmu0}-\ref{fig:lfvtau2}, the bounds derived
in~\cite{DAVIDSON} have been updated to take into account the latest 
results on suppressed or forbidden decays~\cite{PDG98},
and on the $\mu-e$ nuclear conversion~\cite{SINDRUM2}.
Significant such updates concern in particular
$D$ and $K$ decays into $\mu e$,
$B \rightarrow \mu e$~\cite{CDFBDECAY} as well as 
$K \rightarrow \pi \nu \bar{\nu}$, $\tau \rightarrow \pi e$,
$\tau \rightarrow e \gamma$ and $\mu \rightarrow e \gamma$~\cite{MEGA}.
Previous HERA results~\cite{ZEUSLFV}, now superseeded, are given in cases 
where there exists no known indirect constraint.

%
%
%
%
Provided that the quarks involved do not both belong to the first generation, 
it is seen that in some cases H1 limits supersede or come close to  
existing indirect bounds. 
For LQs coupling to muons, this concerns in particular the leptoquarks which 
can contribute to $\mu \rightarrow e \gamma$ and $D \rightarrow \mu e$
processes. 
For LQs coupling to taus, this concerns more cases and in particular the 
leptoquarks which can contribute to 
$\tau \rightarrow K e$, $\tau \rightarrow e \gamma$ and various forbidden
$B$ decays. The sensitivity improves over indirect constraints by up to
an order of magnitude and some cases are covered uniquely by HERA 
experiments.

\section{Conclusions}
\label{sec:conclu}

First generation leptoquarks (LQs) as well as leptoquarks possessing 
lepton flavor violating couplings have been searched at the HERA collider using
1994 to 1997 H1 data in a mass range extending from $75 \GeV$ 
to beyond the $ep$ kinematic limit of $\sqrt{s_{ep}} \simeq 300 \GeV$.

The search for leptoquarks which only couple to first generation fermions 
involved an analysis of very high $Q^2$ neutral (NC) and charged current 
(CC) deep-inelastic scattering data. The comparison of these data with  
Standard Model expectations has shown deviations in the $Q^2$ spectrum 
at $Q^2 \, \gsim \, 10000 \GeV^2$ which are less significant than those previously 
observed in 1994 to 1996. 
No significant clustering of events in excess of SM expectations has
been found in the mass spectra for NC or CC-like events in the 1997 
dataset alone.
Exclusion domains for LQ masses and couplings have been derived.

For first generation leptoquarks of the Buchm\"uller-R\"uckl-Wyler (BRW) 
effective model, masses up to $275 \GeV$ ($284 \GeV$) are excluded for scalars 
(vectors) with a Yukawa coupling of electromagnetic strength, 
$\lambda = \sqrt{4 \pi \alpha_{EM}} = 0.3$.
Constraints on the LQ couplings have been established for $\lambda \lsim 1.0$ 
for all LQ types for masses up to $400 \GeV$.
For $\lambda = 0.3$ but in generic models with arbitrarily small
decay branching ratio $\beta_e$ into NC-like final states, the exclusion 
domain extends to $260 \GeV$ for $\beta_e$ as small as $10 \%$, 
far beyond the present reach of other existing colliders.

No event candidate has been found with either $\mu + jet$ or $\tau + jet$ 
final states compatible with the production of a LQ with couplings
mixing the first and second or third generation in the leptonic sector.
The constraints derived on the Yukawa couplings extend for some
LQ types and coupling products beyond the reach of other colliders as well 
as of low energy experiments. 

\section*{Acknowledgements}
We are grateful to the HERA machine group whose outstanding efforts
made this experiment possible.
We appreciate the immense effort of the engineers and technicians who
constructed and maintain the detector.
We thank the funding agencies for their financial support of the experiment.
We wish to thank the DESY directorate for the hospitality extended
to the non-DESY members of the Collaboration.
We thank W.~Buchm\"uller, M.~Spira, R.~R\"uckl and F.~Schrempp for usefull
help and discussions. 
 
\vfill
\clearpage
{\Large\normalsize}


\begin{thebibliography}{99}
 
%
%

\bibitem{GUT}
 J.C.~Pati and A.~Salam, Phys. Rev. D10 (1974) 275; 
 P.~Langacker, Phys. Rep. 72 (1981) 185;
 H.~Georgi and S.L.~Glashow, Phys. Rev. Lett. 32 (1974) 438.
\vspace{-2mm}

\bibitem{E6}
 A.~Dobado, M.J.~Herrero and C.~Mu\~noz, Phys. Lett. B191 (1987) 449;
 J.F.~Gunion, E.~Ma, Phys. Lett. B195 (1987) 257;
 R.W.~Robinett, Phys. Rev. D37 (1988) 1321;
 J.A.~Grifols and S.~Peris, Phys. Lett. B201 (1988) 287.
\vspace{-2mm}

\bibitem{COMPOSITE}
 B.~Schrempp and F.~Schrempp, Phys. Lett B153 (1985) 101, 
 {\it and references therein};
 J.~Wudka, Phys. Lett. B167 (1986) 337.
\vspace{-2mm}

\bibitem{TECHNICO}
 S.~Dimopoulos and L.~Susskind, Nucl. Phys. B155 (1979) 237;
 S.~Dimopoulos, Nucl. Phys. B168 (1980) 69;
 E.~Farhi, L.~Susskind, Phys. Rev. D20 (1979) 3404;
 {\it idem} Phys. Rep. 74 (1981) 277.
\vspace{-2mm}

\bibitem{H1HIQ2}
 H1 Collaboration, C.~Adloff {\it et al.}, Z. Phys. C74 (1997) 191. 
\vspace{-2mm}

\bibitem{ZEUSHIQ2}
 ZEUS Collaboration, J.~Breitweg {\it et al.}, Z. Phys. C74 (1997) 207.
\vspace{-2mm}

\bibitem{DREESBERN}
 M.~Drees, Phys. Lett. B403 (1997) 353; 
 U.~Bassler and G.~Bernardi, Z. Phys. C76 (1997) 223. 
\vspace{-2mm}

\bibitem{HIGHXYLQ}
T.K.~Kuo and T.~Lee, Mod. Phys. Lett. A12~(1997)~2367; 
K.S.~Babu {\it et al.},  Phys. Lett. B402~(1997)~367;
J.L.~Hewett and T.G.~Rizzo, Phys. Lett. B403~(1997)~353;
Z. Kunszt and W.J. Stirling, Z. Phys. C75~(1997)~453; 
T. Plehn {\it et al.}, Z.~Phys. C74~(1997)~611;
C. Friberg, E. Norrbin and T. Sj\"ostrand, Phys. Lett. B403 (1997) 329; 
J.K. Elwood and A.E. Faraggi, Nucl. Phys. B512 (1998) 42;
M.~Heyssler and W.J.~Stirling, Phys. Lett. B407 (1997) 259;
J.~Bl\"umlein, Proceed. of the 5th Int. Workshop on Deep Inelastic Scattering
               and QCD (DIS 97), Chicago, USA (14-18 April 1997) 5pp.; 
E.~Keith and E.~Ma, Phys. Rev. Lett. 79 (1997) 4318;
N.G.~Deshpande and B.~Dutta, Phys. Lett. B424 (1998) 313;
J.L.~Hewett and T.G. Rizzo, SLAC preprint PUB-7549 (August 1997) 45pp.;  
T.G.~Rizzo, Proceed. of the Workshop on Physics Beyond the Desert 1997,
             Tegernsee, Germany (8-14 June 1997) 32pp.;
Z.~Xiao, RAL preprint TR-97-043 (September 1997) 26pp.;
R.~R\"uckl and H.~Spiesberger, Proceed. of the Workshop on Physics Beyond the 
             Desert, Tegernsee, Germany (8-14 Jun 1997) 18pp.;
M.~Sekiguchi, H.~Wada and S.~Ishida, Nihon Univ. preprint NUP-A-97-23
             (December 1997) 7 pp.
\vspace{-2mm}

\bibitem{BUCH1987}
 W.~Buchm\"uller, R.~R\"uckl and D.~Wyler, Phys.~Lett.~B191 (1987) 442.
 {\it{Erratum}} Phys.~Lett.~B448 (1999) 320.
\vspace{-2mm}

\bibitem{H1F2PAPER}
 H1 Collaboration, C.~Adloff {\it et al.},
 ``Measurement of Neutral and Charged Current cross-sections
   in Positron-Proton Collisions at Large Momentum Transfer'',
   {\it To be submitted for publication}.
\vspace{-2mm}

\bibitem{H1LQ}
 H1 Collaboration, I.~Abt {\it et al.}, Nucl. Phys. B396 (1993) 3;
       {\it idem}, T.~Ahmed {\it et al.}, Z.~Phys.~C64 (1994) 545; 
       {\it idem}, T.~Ahmed {\it et al.}, Phys. Lett. B369 (1996) 173.
\vspace{-2mm}
 
\bibitem{ZEUSLQ}
 ZEUS Collaboration, M.~Derrick {\it et al.}, Phys. Lett. B306 (1993) 173.
\vspace{-2mm}

\bibitem{ZEUSLFV}
 ZEUS Collaboration, M.~Derrick {\it et al.}, Z. Phys. C73 (1997) 613.
\vspace{-2mm}

%
%

\bibitem{H1DETECT}
 H1 Collaboration, I.~Abt {\it et al.}, 
 Nucl. Instr. and Meth. A386 (1997) 310.;
 {\it idem} Nucl. Instr. and Meth. A386 (1997) 348.;
 H1 Spacal Group, R.~Appuhn {\it et al.},
 Nucl. Instr. and Meth. A386 (1997) 397.
\vspace{-2mm}

\bibitem{H1LARCAL}
 H1 Calorimeter Group, B.~Andrieu {\it et al.},
 Nucl. Instr. and. Meth. A336 (1993) 460.
\vspace{-2mm}

\bibitem{H1CALEPI}
 H1 Calorimeter Group, B. Andrieu {\it et al.},
 Nucl. Inst. and Meth. A344 (1994) 492.
\vspace{-2mm}

\bibitem{H1CALRES}
 H1 Calorimeter Group, B.~Andrieu {\it et al.},
 Nucl. Instr. and. Meth. A350 (1994) 57;
{\it idem}, Nucl. Instr. and. Meth. A336 (1993) 499.
\vspace{-2mm}

\bibitem{PHDBRUEL}
 Ph.~Bruel, Ph.D Thesis, Universit\'e de Paris-Sud,
 ``Recherche d'int\'eractions au-del\`a du Mod\`ele Standard''
 (in French) (1998).
\vspace{-2mm}

\bibitem{H1SPACAL}
 H1 SPACAL Group, R.D. Appuhn {\it et al.}, 
 Nucl. Instr. and Meth. A386 (1997) 397. 
\vspace{-2mm}

\bibitem{H1BEMC}
 H1 BEMC Group, J. Ban {\it et al.}, Nucl. Instr. and Meth. A372 (1996) 399. 
\vspace{-2mm}


%
%
 
\bibitem{DURHAM}
 T.~Matsushita, E.~Perez and R.~R\"uckl, Contribution to the 3rd UK 
 Phenomenology Workshop on HERA Physics, St. John's College, Durham,
 UK, September 1998, hep-ph/9812481, WUE-ITP-98-056,
 DAPNIA/SPP 98-23, 15pp.
\vspace{-2mm}

\bibitem{DAVIDSON}
S.~Davidson, D.~Bailey and B.~Campbell, Z. Phys. C61 (1994) 613.
\vspace{-2mm}
 
\bibitem{LEURER}
M.~Leurer, Phys. Rev. D49 (1994) 333; {\it idem.} D50 (1994) 536.
\vspace{-2mm}

\bibitem{HERAWS}
B.~Schrempp, Proc. of the Workshop Physics at HERA,
DESY, Hamburg (1991), vol. 2 p. 1034, {\it and references therein}.
\vspace{-2mm}

%
%

\bibitem{LEGO}
 LEGO~0.02 and SUSSEX~1.5;
 K.~Rosenbauer, dissertation RWTH Aachen (in German), PITHA 95/16, July 1995.
\vspace{-2mm}

\bibitem{JETSET74}
 JETSET~7.3 and 7.4;
 T.~Sj\"ostrand, Lund Univ. preprint LU-TP-95-20 (August 1995) 321pp;
 {\it idem}, CERN preprint TH-7112-93 (February 1994) 305pp.
\vspace{-2mm}

\bibitem{DGLAP}
 V.N. Gribov et L.N. Lipatov, Sov. Journ. Nucl. Phys. 15 (1972) 78;\\
 G. Altarelli et G. Parisi, Nucl. Phys. B126 (1977) 298;\\
 Y.L. Dokshitzer, JETP 46 (1977) 641.
\vspace{-2mm}

\bibitem{SPIRA}
 T.~Plehn {\it{et al.}}, Z. Phys. C74 (1997) 611.
\vspace{-2mm}

\bibitem{MRSTSF}
 A.D.~Martin, R.G.~Roberts, W.J.~Stirling and R.S.~Thorne,
 Euro. Phys. J. C4 (1998) 463.
\vspace{-2mm}

\bibitem{NA51}
 NA51 Collaboration, A.~Baldit {\it{et al.}},
 Phys. Lett. B332 (1994) 244.
\vspace{-2mm}

\bibitem{DRELLYAN}
 E866 Collaboration, E.A.~Hawker {\it{et al.}}, 
 Phys. Rev. Lett. 80 (1998) 3715. 
\vspace{-2mm}

\bibitem{H1MRST}
 H1 Collaboration, S.~Aid {\it{et al.}}, 
 Nucl. Phys. B470 (1996) 3.
\vspace{-2mm}

\bibitem{ZEUSMRST}
 ZEUS Collaboration, M.~Derrick {\it{et al.}},
 Zeit. Phys. C72 (1996) 399.
\vspace{-2mm}

\bibitem{DJANGO} 
 DJANGO~6.2;
 G.A.~Schuler and H.~Spiesberger,
 Proc. of the Workshop Physics at HERA,
 W.~Buchm\"uller and G.~Ingelman (Editors),
 (October 1991, DESY-Hamburg) Vol. 3 p. 1419.
\vspace{-2mm}

\bibitem{HERACLES}
 HERACLES 4.4;
 A.~Kwiatkowski, H.~Spiesberger and H.-J.~M\"ohring,
 Comput.~Phys.~Commun. 69 (1992) 155.
\vspace{-2mm}

\bibitem{ARIADNE}
 ARIADNE 4.08;
 L.~L\"onnblad, Comput.~Phys.~Commun. 71 (1992) 15.
\vspace{-2mm}

\bibitem{CDM}
 G. Gustafson and U. Pettersson, Nucl. Phys. B306 (1988) 746;
 {\it idem}, {\it addendum} Lund University preprint LU-TP-87-19,
 (October 1987) 4pp.;
 B.~Andersson {\it et al.}, Z. Phys. C43 (1989) 625. 
\vspace{-2mm}

\bibitem{PYTHIA}
 PYTHIA~5.6;
 T.~Sj\"ostrand, Comp. Phys. Comm. 39 (1986) 347; \\
 T.~Sj\"ostrand and M.~Bengtsson, Comp. Phys. Comm. 43 (1987) 367.
\vspace{-2mm}
 
\bibitem{GRVG}
 M.~Gl\"uck, E.~Reya and A.~Vogt, Phys. Rev. D45 (1992) 3986;
 {\it idem}, Phys. Rev. D46 (1992) 1973.
\vspace{-2mm}

\bibitem{EPVEC}
 H1 generator based on EPVEC 1.0;
 U.~Baur, J.A.M.~Vermaseren and D.~Zeppenfeld, Nucl. Phys. B375 (1992) 3.
\vspace{-2mm}

\bibitem{LPAIR}
 S.~Baranov {\it{et al.}}, Proc. of the Workshop Physics at HERA,
 W.~Buchm\"uller and G.~Ingelman (Editors),
 (October 1991, DESY-Hamburg) Vol. 3, p. 1478;
 J.A.M.~Vermaseren, Nucl. Phys. B229 (1983) 347.
\vspace{-2mm}

%
%

\bibitem{HOEGER}
 S. Bentvelsen, J. Engelen and P. Kooijman, 
 Proc. of the Workshop Physics at HERA,
 W.~Buchm\"uller and G.~Ingelman (Editors),
 (October 1991, DESY-Hamburg) Vol. 1 p. 25;
 K.C. Hoeger, {\it idem.} p. 43; {\it and references therein}.
\vspace{-2mm}

\bibitem{JACQUET}
 A.~Blondel, F.~Jacquet, Proceedings of the Study of an
 $ep$ Facility for Europe, ed. U.~Amaldi,
 DESY report 79-48 (1979) 391.
\vspace{-2mm}

\bibitem{H1MUEV}
 H1 Collaboration, C.~Adloff {\it et al.}, 
 Euro. Phys. J. C5 (1998) 575.
\vspace{-2mm}

%
%
 
\bibitem{DELPHI} 
 DELPHI Collaboration, P.~Abreu {\it et al.},
 Phys. Lett. B446 (1999) 62. 
\vspace{-2mm}

\bibitem{ALEPH}
 ALEPH Collaboration,
 R.~Barate {\it et al.},
 ``Study of Fermion Pair Production in $e^+ e^-$ Collisions
   at $130 - 183 \GeV$'',
 CERN-EP/99-042, Apr. 1999
\vspace{-2mm}

\bibitem{OPAL}
 OPAL Collaboration, G.~Abbiendi {\it et al.},
 Euro. Phys. J. C6 (1999) 1.
\vspace{-2mm}

\bibitem{L3}
 L3 Collaboration, M.~Acciarri {\it et al.},
 Phys. Lett. B433 (1998) 163. 
\vspace{-2mm}
 
\bibitem{D01GENE}
 D$\emptyset$ Collaboration, B. Abbott {\it et al.}, 
 Phys. Rev. Lett. 79 (1997) 4321;\\
 D$\emptyset$ Collaboration, B. Abbott {\it et al.}, 
 Phys. Rev. Lett. 80 (1998) 2051.
\vspace{-2mm}

\bibitem{CDF1GENE}
 CDF Collaboration, F. Abe {\it et al.}, Phys. Rev. Lett. 79 (1997) 4327.
\vspace{-2mm}

\bibitem{TEVCOMBINED}
 LQ Limit Combination Working Group, for the CDF and D$\emptyset$
 Collaborations, 
 ``Combined Limits on First Generation Leptoquarks from the CDF and D0
   Experiments", 
 hep-ex/9810015 (Oct. 98), 9pp.
\vspace{-2mm}

\bibitem{CDF3GENE}
 CDF Collaboration, F. Abe {\it et al.}, Phys. Rev. Lett. 78 (1997) 2906.
\vspace{-2mm}

\bibitem{D03GENE}
 D$\emptyset$ Collaboration, B. Abbott {\it et al.}, 
 Phys. Rev. Lett. 81 (1998) 38. 
\vspace{-2mm}

\bibitem{PDG98}
 Review of Particle Physics,
 European Phys. Journal C, Vol 3, Nb 1-4 (1998).
\vspace{-2mm}

\bibitem{SPIRAEB}
 M.~Spira, Private Communication; see also
 A.~Djouadi {\it et al.}, Z. Phys. C46 (1990) 679.
\vspace{-2mm}

\bibitem{SINDRUM2}
 SINDRUM II Collaboration,
 S.~Eggli {\it et al.},
 ``Search for $\mu^- \to e^-$ Conversion on Titanium",
 submitted to Phys. Rev. C (1999).
\vspace{-2mm}

\bibitem{CDFBDECAY}
 CDF Collaboration, F. Abe {\it et al.},
 Phys. Rev. Lett. 81 (1998) 5742.
\vspace{-2mm}

\bibitem{MEGA}
 MEGA Collaboration, M.L.~Brooks {\it et al.},
 LA-UR-99-2268, hep-ex/9905013.

%
\end{thebibliography}
\end{document}